\documentclass[a4paper,11pt]{article}
\pdfoutput=1

\usepackage{jcappub}
\usepackage[utf8]{inputenc}
\usepackage{arydshln} 
\usepackage{xcolor}
\usepackage{float}
\usepackage{soul}

\renewcommand\eqref[1]{Eq.~(\ref{#1})}

\newcommand\figref[1]{Fig.~\ref{#1}}

\newcommand{\nn}{\nonumber}

\newcommand{\be}{\begin{equation}}
\newcommand{\ee}{\end{equation}}
\newcommand{\bear}{\begin{eqnarray}}
\newcommand{\eear}{\end{eqnarray}}
\newcommand{\mL}{\mathcal{L}}
\newcommand{\mO}{\mathcal{O}}

\newcommand{\Frac}[2]{\frac{\displaystyle #1}{\displaystyle #2}}

\title{Unitarization effects in EFT predictions of  WZ scattering at the LHC}
\author[a]{Claudia Garcia-Garcia,}
\author[a]{Maria Herrero,}
\author[a]{and\\ Roberto A. Morales}
\affiliation[a]{Departamento de F\'{\i}sica Te\'orica and Instituto de F\'{\i}sica Te\'orica, IFT-UAM/CSIC,\\
Universidad Aut\'onoma de Madrid, Cantoblanco, 28049 Madrid, Spain}
\emailAdd{claudia.garcia@uam.es}
\emailAdd{maria.herrero@uam.es}
\emailAdd{robertoa.morales@uam.es}

\abstract{
	Effective field theories are an incredibly powerful tool in order to study and understand the true nature of the 
symmetry breaking sector dynamics of the Standard Model. However, they can suffer from some theoretical problems such as that of unitarity violation. Nevertheless, in order to interpret experimental data correctly a fully unitary prescription is needed. To this purpose, unitarization methods are addressed, but each of them leads to a different (unitary) prediction. Because of this, there is an inherent theoretical uncertainty in the determination of the effective field theory parameters due to the choice of one unitarization scheme. In this work, we quantify this uncertainty assuming a strongly interacting electroweak symmetry breaking sector, described by the effective electroweak chiral Lagrangian. We focus on the bosonic part of this effective Lagrangian and choose in particular the  WZ scattering as our main VBS channel to study the sensitivity to new physics at the LHC. We study the different predictions of various well known unitarization methods, considering the full coupled system of helicity amplitudes, and construct the 95\% confidence level exclusion regions for  the most relevant electroweak chiral Lagrangian parameters, given by the two anomalous quartic gauge couplings $a_4$ and $a_5$. This provides a consistent analysis of the different constraints on EChL parameters that can be achieved by using different unitarization methods in a combined way.}

\begin{document}
\begin{flushright}
	IFT-UAM/CSIC-19-81  \\
	FTUAM-19-13
\end{flushright}
\maketitle


\section{Introduction}

Although the discovery of the Higgs boson by the ATLAS and CMS experiments supposed a great success of the Standard Model (SM), it also posed a lot of new questions about the symmetry breaking sector of  the Electroweak (EW) Theory. Questions such as why is the Higgs boson so light, being its mass so similar to that of the EW gauge bosons, is the Higgs boson an elementary or a composite particle, what mechanism generates its potential, and others. All this can be summarized in the fact that the dynamical generation of electroweak symmetry breaking (EWSB) is still a mystery to be solved. 

A very efficient way to try to understand the true nature of the EWSB  sector of the SM is to use effective field theories (EFTs). Effective theories allow us to describe in a model independent way the relevant beyond the SM (BSM) physics that might be responsible for the dynamical generation of EWSB. In these theories the ultraviolet (UV) dynamics are, in principle, unknown, but their effects at low energies remain present encoded in a finite set of low energy parameters. If these low energy parameters were measured, we would have a hint towards the UV completion that might describe best the true dynamics of the EWSB sector. 

With the aim of obtaining such a measurement, many experimental searches at the LHC are devoted to look for signals predicted by these effective theories. The most characteristic of these signals are those coming from vector boson scattering (VBS) processes (for recent reviews on VBS physics see, for instance, \cite{Rauch:2016pai,deFlorian:2016spz, Anders:2018gfr,Bellan:2019xpr} and references therein) since it is there where the interactions among  the longitudinal EW gauge bosons will appear dominantly, due to their relation to the scalar Goldstone boson modes which are associated to the spontaneous symmetry breaking taking place in the EW Theory. 

Nevertheless, predictions of observable rates computed with effective theories can carry some theoretical problems.  It is typical from such theories, due to the energy structure of the operators involved, to suffer from unitarity violation problems at high energies, such as the ones being now probed by the LHC. However, predictions that are to be tested at colliders must be fully unitary to be consistent with the underlying quantum field theory. Therefore, a prescription is needed to translate these non-unitary predictions into reliable, unitary ones with which interpret the experimental data. These prescriptions are called unitarization methods or procedures, that drive unitary the non-unitary EFT predictions (for some illustrative reviews on different unitarization methods in the context of VBS see, for instance, \cite{Kilian:2014zja, Rauch:2016pai,DelgadoLopez:2017ugq}). The problem with these methods is that the various manners of unitarizing the computation of an observable lead to different final results \cite{Alboteanu:2008my,Espriu:2012ih,Delgado:2013loa,Espriu:2014jya,Kilian:2014zja,Delgado:2015kxa,Corbett:2014ora,Corbett:2015lfa,Rauch:2016pai,DelgadoLopez:2017ugq,Perez:2018kav,Kozow:2019txg}. Thus, a theoretical uncertainty arises when computing unitarized EFT  predictions due to the fact that there is a variety of ways of achieving such a unitary outcome.  

Current constraints imposed on some of the mentioned low-energy parameters by LHC experiments do not take this theoretical uncertainty into account. They are all basically based on searches for anomalous quartic gauge couplings that are then interpreted using the theoretical EFT predictions in different ways, i.e., using different unitarization methods or no unitarization method at all. 
For instance, the most recent constraints given in 
\cite{Aaboud:2016uuk, Sirunyan:2019der,Aad:2019xxo,Sirunyan:2019ksz} provide a model independent experimental analysis, do not rely upon any particular method or employ one unitarization method only.

In this work, we quantify the uncertainty due to the choice of unitarization scheme present in the determination of some of the most relevant low-energy constants for VBS processes. To this aim, we assume a strongly interacting EWSB sector, properly described by the effective electroweak chiral Lagrangian (EChL) 
(nowdays called also Higgs effective field theory (HEFT)) 
(see, for instance, \cite{Appelquist:1980vg,Longhitano:1980iz,Chanowitz:1985hj,Cheyette:1987jf, Dobado:1989ax,Dobado:1989ue,Dobado:1990jy,Dobado:1990am,Dobado:1990zh,
Espriu:1991vm,Feruglio:1992wf,Dobado:1995qy,Dobado:1999xb,Eboli:2006wa}, before the Higgs boson discovery, and \cite{Alonso:2012px,
Buchalla:2013rka,Espriu:2012ih, Delgado:2013loa,Delgado:2013hxa,Brivio:2013pma,Espriu:2013fia,Espriu:2014jya,
Delgado:2014jda,Delgado:2015kxa,Corbett:2014ora,Corbett:2015lfa,Buchalla:2015qju,Buchalla:2017jlu,Delgado:2017cls,Perez:2018kav,Kozow:2019txg} after this discovery)
and we focus on the bosonic sector of this Lagrangian, choosing to study the particular VBS process given by the WZ channel, as an example which is also interesting from the experimental detection perspective. Within this framework, we characterize the unitarity violation that arises in the predictions of the WZ $\to$ WZ cross sections, and we analyze the impact that a variety of well stablished unitarization methods have on them. We pay special attention to the fact that all helicity states of the incoming and outgoing gauge bosons might play a relevant role in the unitarization process, and consider them all at once as a coupled system. Then, we move on to the LHC scenario.  We use the Effective W Approximation (EWA)~\cite{Dawson:1984gx,Johnson:1987tj} to give predictions of $pp\to$WZ$+X$ events at the LHC for different unitarization schemes. In order to check that the EWA works for our purpose here, we compare the EWA predictions for the cases of the SM and the EChL with the corresponding  full results from the Monte Carlo MadGraph~v5 (MG5)~\cite{Alwall:2014hca,Frederix:2018nkq}, and we find very good agreement in both cases.  Finally, in order to provide a quantitative analysis of the implications of our study on the LHC searches, we choose in particular to compare our results with those in  \cite{Aaboud:2016uuk}. Concretely, 
we translate the ATLAS constraints from \cite{Aaboud:2016uuk} to construct the 95\% exclusion regions in some of the EChL parameter space for each of the considered unitarization methods, giving, at the same time, the total theoretical uncertainty driven by the  variety of these methods, which represents our main conclussions in this work.

The paper is organized as follows. In section~\ref{EChL} we summarize the main features
of the EChL and the relevant operators for describing deviations with respect to the SM in VBS processes. We also  introduce the issue of unitarity violation in effective field theories. In section~\ref{Unitarity} we present the different methods we will use to deal with this problem and the corresponding predictions for the 
WZ $\to$ WZ scattering process within the EChL framework. We will also comment on the importance of taking into account the whole coupled system of helicity amplitudes, contrary to what is usually done in the literature.
 Section~\ref{LHC} is devoted to the presentation of our main results, in which we show the different predictions of the various unitarization methods in VBS processes at the LHC. In this section we display the impact of using different unitarization methods on the constraints that can be imposed on the relevant EChL parameters.
The final section summarizes our main conclusions.



\section{The Electroweak Chiral Lagrangian and the violation of unitarity}
\label{EChL}

As already introduced in the previous section, we will work under the assumption of a strongly interacting EWSB sector. Even though the description of physics beyond the SM is unavoidably model dependent, we will employ a technology that is as model independent as possible. This technology is based on the effective theory parametrization of the possible BSM interactions, and the appropriate approach to a strongly interacting EWSB sector in this framework is to use the effective EChL, also known as HEFT in the literature. In this context, the information of the short-range theory is encoded in a certain number of coefficients of local operators, often called low-energy parameters.

The EChL is a gauged non-linear effective field theory based on the $SU(2)_L \times SU(2)_R$ chiral symmetry of the EWSB sector that is spontaneously broken down to the subgroup $SU(2)_{L+R}$, usually called the custodial symmetry group or EW isospin group. It contains the EW gauge bosons, $W^{\pm}$, $Z$ and $\gamma$, their corresponding would-be Goldstone bosons, $w^{\pm}$, $z$, and the Higgs scalar boson, $H$ as dynamical fields, being the latter a singlet under the EW chiral symmetry. For the sake of simplicity, and since their contribution to VBS processes in the strongly interacting EWSB case should be negligible, we will not discuss the fermion sector in this article. 

The $w^{\pm}$, $z$ are introduced in a matrix field $U(w^\pm,z) = 1 + i w^a \tau^a/v + \mO(w^2)$ that takes values in the $SU(2)_L \times SU(2)_R/SU(2)_{L+R}$ coset, and transforms as $U \to g_L U g_R^\dagger$ under the $SU(2)_L \times SU(2)_R$  global group. The EW gauge bosons are introduced through the following covariant derivative and field strength tensors:
 \begin{align}
D_\mu U &= \partial_\mu U + i\hat{W}_\mu U - i U\hat{B}_\mu \,, \label{eq.cov-deriv} \\
\hat{W}_{\mu\nu} &= \partial_\mu \hat{W}_\nu - \partial_\nu \hat{W}_\mu
 + i  [\hat{W}_\mu,\hat{W}_\nu ],\;\hat{B}_{\mu\nu} = \partial_\mu \hat{B}_\nu -\partial_\nu \hat{B}_\mu \,,
\label{fieldstrength}
\end{align}
with $\hat{W}_\mu = g \vec{W}_\mu \vec{\tau}/2 ,\;\hat{B}_\mu = g'\, B_\mu \tau^3/2$.
Finally, the Higgs boson, being it a singlet of the EW chiral symmetry, is described by a generic polynomial function ${\cal F}(H)= 1+2a\frac{H}{v}+b \left(\frac{H}{v} \right)^2+\dots$, where the parameters $a$ and $b$, if different from 1, describe the new BSM interactions of the Higgs boson to the EW gauge bosons.

Here we will asume that custodial symmetry is preserved in the EWSB sector, except for the explicit breaking due to the gauging of the $U(1)_Y$ symmetry. This assumption is based on experimental measurements of the $\rho$ parameter and of the effective couplings between the Higgs and the EW gauge bosons, that disfavor custodial breaking other than that induced from $g'\neq0$. 

According to the usual counting rules within the chiral Lagrangian approach \footnote{Throughout this work, we will use the same conventions and counting rules presented in \cite{Delgado:2014jda}.},  the different EChL operators are organized by means of their `chiral dimension'. This chiral dimension can be found by following the scaling with the external momentum, $p$, of the various contributing basic functions, since, after all, the EChL structure is based on a momentum expansion. Derivatives and masses are considered as soft scales of the EFT and of the same order in the chiral counting, i.e., of ${\cal O}(p)$. Furthermore, in order to have a power counting consistent with
the loop expansion, one needs $\partial_\mu$, $(g v)$ and  $(g'v) \sim \mO(p)$ or, equivalently,
$\partial_\mu$, $m_W$,  $m_Z \sim \mO(p) $. The typical energy scale that controls the size of the various contributing terms in this chiral expansion is provided by 
$4 \pi v$, where $v=246$ GeV  is the vacuum expectation value of the Higgs field, in close analogy to the typical scale of $4 \pi f_\pi$, with $f_\pi=94$ MeV, for the case of the chiral Lagrangian in QCD. In the scenarios where there are resonances that emerge typically from the assumed strongly interacting underlying UV theory, then there are additional mass scales given by the masses of the resonances to account for in the EChL. However, in this work we will assume that there are not emergent resonances below roughly $4 \pi v \sim 3$ TeV and, therefore, this will be our unique energy scale parameter in the EW chiral expansion. All the other masses involved, $m_H$, $m_W$ and $m_Z$, are soft masses, as we have already said.     

With all these considerations in mind, one then constructs the EChL up to a given order in the chiral
expansion. This Lagrangian must be $CP$, Lorentz and $SU(2)_L \times U(1)_Y$ gauge invariant. Furthermore, with our simplifying assumption, it should be as well custodial symmetry preserving. For the present work, we include the terms with
chiral dimension up to  $\mO(p^4)$, thus, the EChL can be generically written as:
\be
\mL_{\rm EChL} = \mL_2 + \mL_4 +\mL_{\rm GF} +\mL_{\rm FP}\, ,
\ee
where  $\mL_2$ refers to the terms with chiral dimension $\mO(p^2)$,  $\mL_4$ refers
to the terms with chiral dimension $\mO(p^4)$, and
$\mL_{\rm GF}$ and $\mL_{\rm FP}$ are the gauge-fixing (GF) and the corresponding Fadeev-Popov (FP) terms.  

Nevertheless, not all the operators that can be included a priori in $\mL_2$ and $\mL_4$ have the same relevance in VBS processes. We conclude that for our purpose of describing the most relevant deviations from the SM in VBS it will be sufficient to work with just a subset of EChL operators, i.e., the most relevant ones. These operators are the following\footnote{Again, we use the same conventions for the EChL effective operators  as in \cite{Delgado:2014jda}.}: 
\begin{align}
\mL_2 =&    -\Frac{1}{2 g^2} {\rm Tr}\Big(\hat{W}_{\mu\nu}\hat{W}^{\mu\nu}\Big) -\Frac{1}{2 g^{'2}}
{\rm Tr} \Big(\hat{B}_{\mu\nu} \hat{B}^{\mu\nu}\Big)\nn\\
& +\Frac{v^2}{4}\left[%
  1 + 2a \frac{H}{v} + b \frac{H^2}{v^2}\right] {\rm Tr} \Big(D^\mu U^\dagger D_\mu U \Big)
 + \Frac{1}{2} \partial^\mu H \, \partial_\mu H + \dots\, ,
\label{eq.L2}
\end{align}
\begin{align}
{\mL}_{4} =& %
 ~ a_1 {\rm Tr}\Big( U \hat{B}_{\mu\nu} U^\dagger \hat{W}^{\mu\nu}\Big)
  + i a_2  {\rm Tr}\Big ( U \hat{B}_{\mu\nu} U^\dagger [{\cal V}^\mu, {\cal V}^\nu ]\Big)
  - i a_3 {\rm Tr}\Big (\hat{W}_{\mu\nu}[{\cal V}^\mu, {\cal V}^\nu]\Big) \nn \\
+&
  ~ a_4 \Big[{\rm Tr}({\cal V}_\mu {\cal V}_\nu) \Big]  \Big[{\rm Tr}({\cal V}^\mu {\cal V}^\nu)\Big] 
  + a_5  \Big[{\rm Tr}({\cal V}_\mu {\cal V}^\mu)\Big]  \Big[{\rm Tr}({\cal V}_\nu {\cal V}^\nu)\Big]  +\dots\label{eq.L4}
\end{align}
where ${\cal V}_\mu = (D_\mu U) U^\dagger$.

In this framework, as we have said, the chiral parameters $a$ and $b$ intervene in the interactions between two EW gauge bosons and one or two Higgs bosons, respectively. The other parameters, the ones controlling the strength of the $\mO(p^4)$ operators, appear in the self interactions of the EW gauge bosons. The SM prediction is recovered for $a=b=1$ and $a_i=0$.

The fact that in the context of this strongly interacting dynamics, operators, and thus, interactions among gauge bosons, scale directly with the external momentum, leads to a scenario in which predictions of observables can behave pathologically with energy from a certain energy scale upwards. This pathology translates into a violation of unitarity of the $S$ matrix, which basically implies an unphysical leak in the interaction probability among EW gauge bosons. The energy at which this violation of unitarity occurs can be easily computed by studying the unitarization condition, implemented at the level of the partial waves, defined in this work as:
\begin{align}
a^J_{\lambda_1\lambda_2\lambda_3\lambda_4}(s)&=\dfrac{1}{32\pi}\int_{-1}^1 d\cos\theta~ A(W_{\lambda_1}Z_{\lambda_2}\to W_{\lambda_3}Z_{\lambda_4})(s,\cos\theta)\, d^J_{\lambda,\lambda'}(\cos\theta) \,, \label{PW}
\end{align}
where $J$ is the total angular momentum of the system, $\lambda=\lambda_1-\lambda_2$, $\lambda'=\lambda_3-\lambda_4$, being $\lambda_i$ the helicity states of the external gauge bosons, and where $d^J_{\lambda,\lambda'}(\cos\theta)$ are the Wigner functions. 

In this framework, the violation of unitarity occurs when the expression:
\begin{equation}
{\rm Im} \Big[a^J_{\lambda_1\lambda_2\lambda_3\lambda_4}(s)\Big]=|a^J_{\lambda_1\lambda_2\lambda_3\lambda_4}(s)|^2=\sum_{\lambda_a,\lambda_b} a^J_{\lambda_1\lambda_2\lambda_a\lambda_b}(s) a^{J*}_{\lambda_a\lambda_b\lambda_3\lambda_4}(s) \,,
\label{unitarity}
\end{equation}
is not fulfilled. Therefore, the violation of unitarity can take place for any value of the angular momentum $J$, and for every helicity channel, in principle.  Furthermore, this expression implies that the unitarity condition of a particular helicity amplitude might depend on the amplitudes corresponding to other helicity channels too. This is an important statement and, therefore, when studying the possible unitarity violation of the various channels involved in WZ scattering, all helicity states should be considered consistently as a coupled system. 

\eqref{unitarity} can be rewritten in a more friendly way in the following manner:
\begin{align}
|a_J(s)|&\leq1\,.\label{unitlim}
\end{align}
This way, the value of the energy at which the $J^{th}$ partial wave crosses the unitarity limit, i.e., $|a_J(s)|=1$, defines the unitarity violation scale.

\begin{figure}[t!]
\begin{center}
\includegraphics[width=0.49\textwidth]{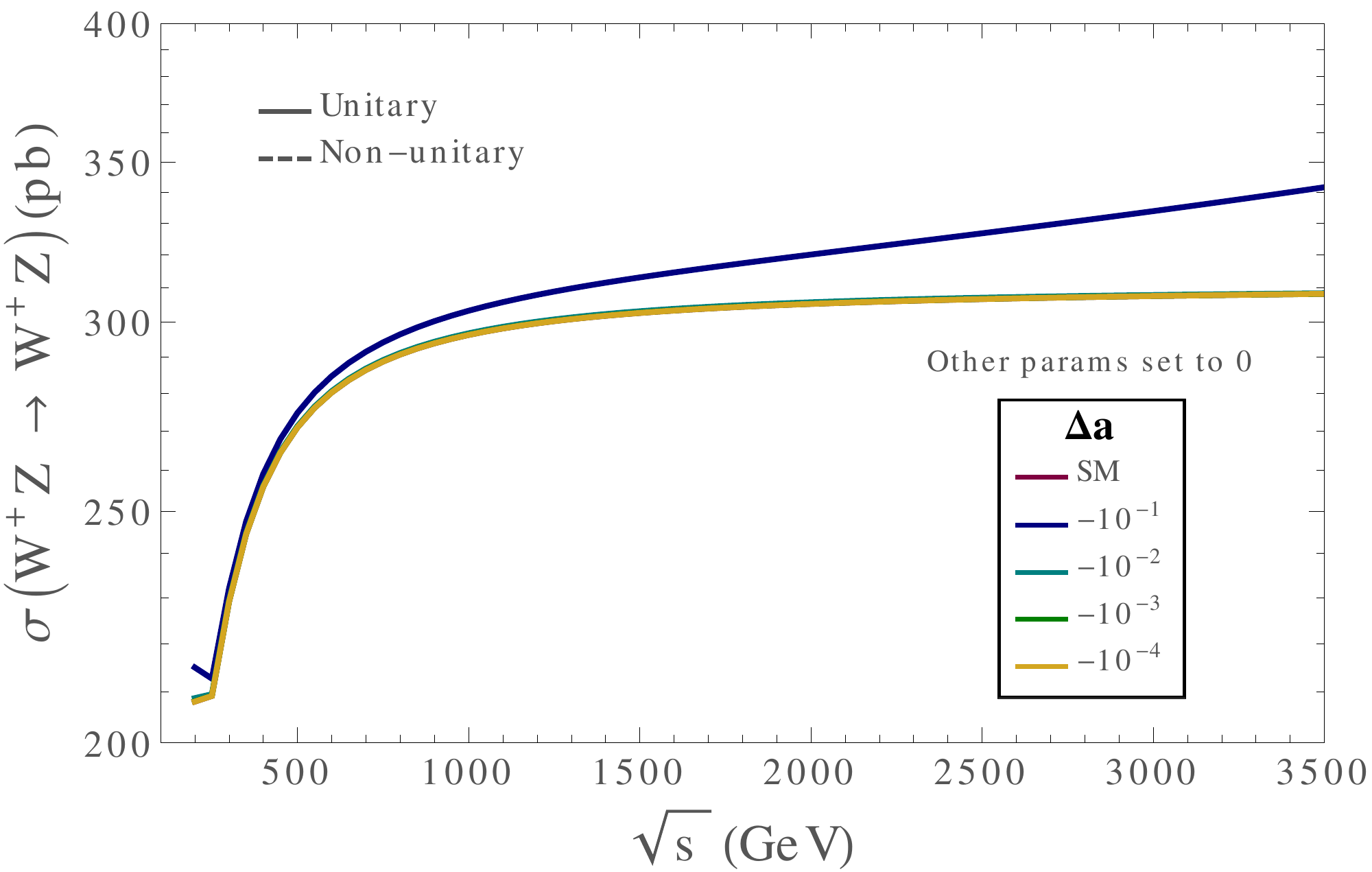}
\includegraphics[width=0.49\textwidth]{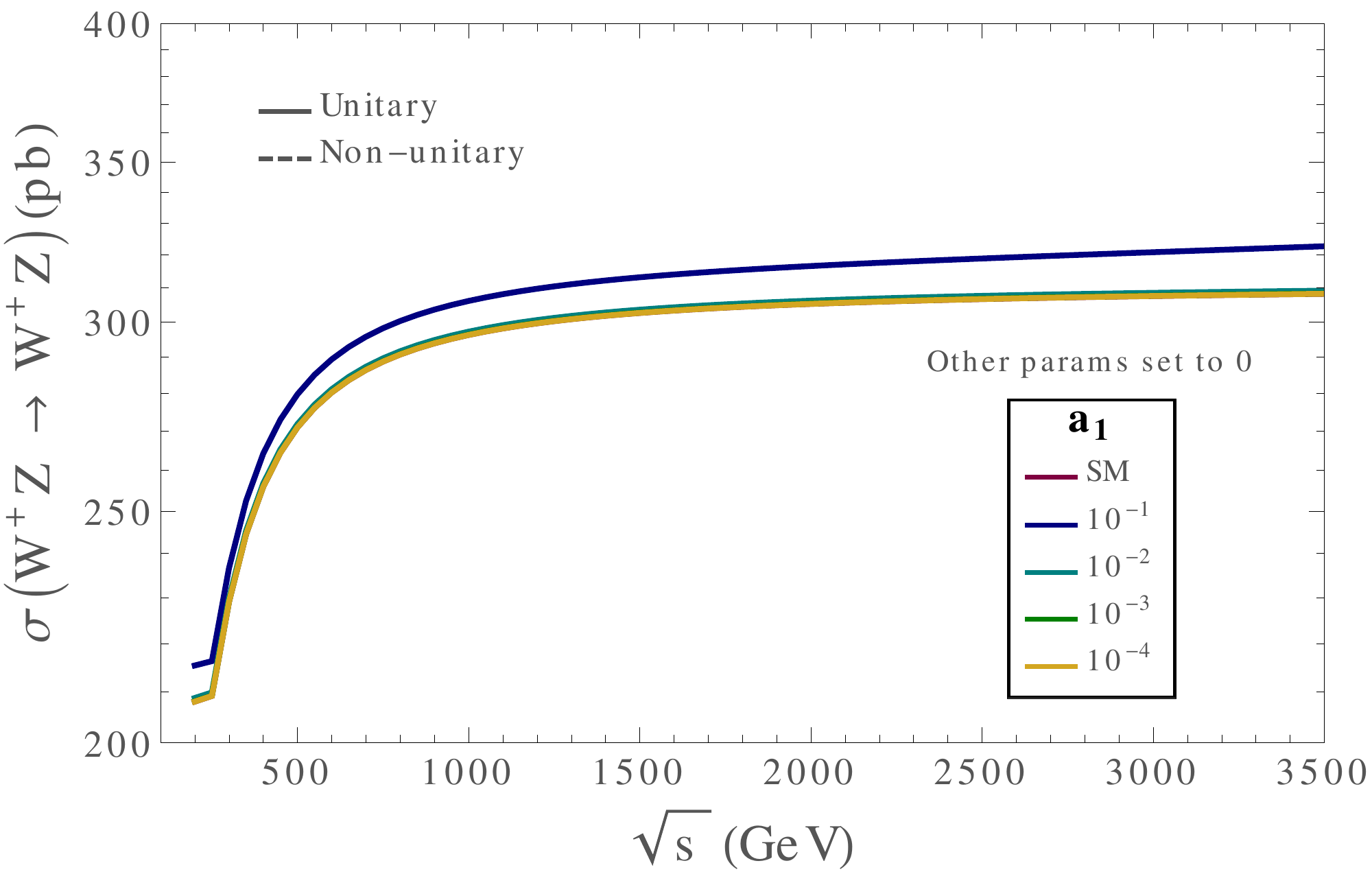}\\
\includegraphics[width=0.49\textwidth]{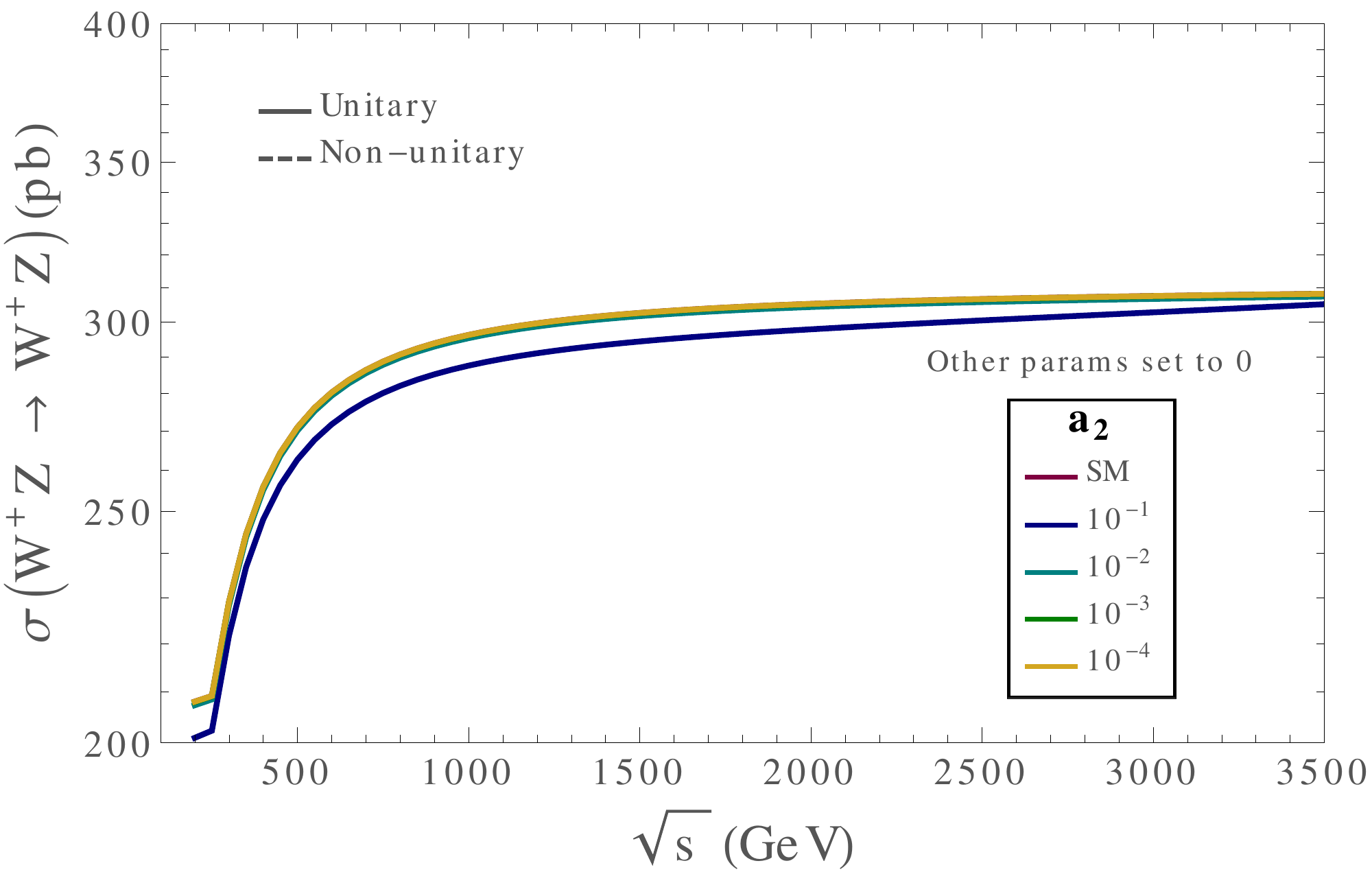}
\includegraphics[width=0.49\textwidth]{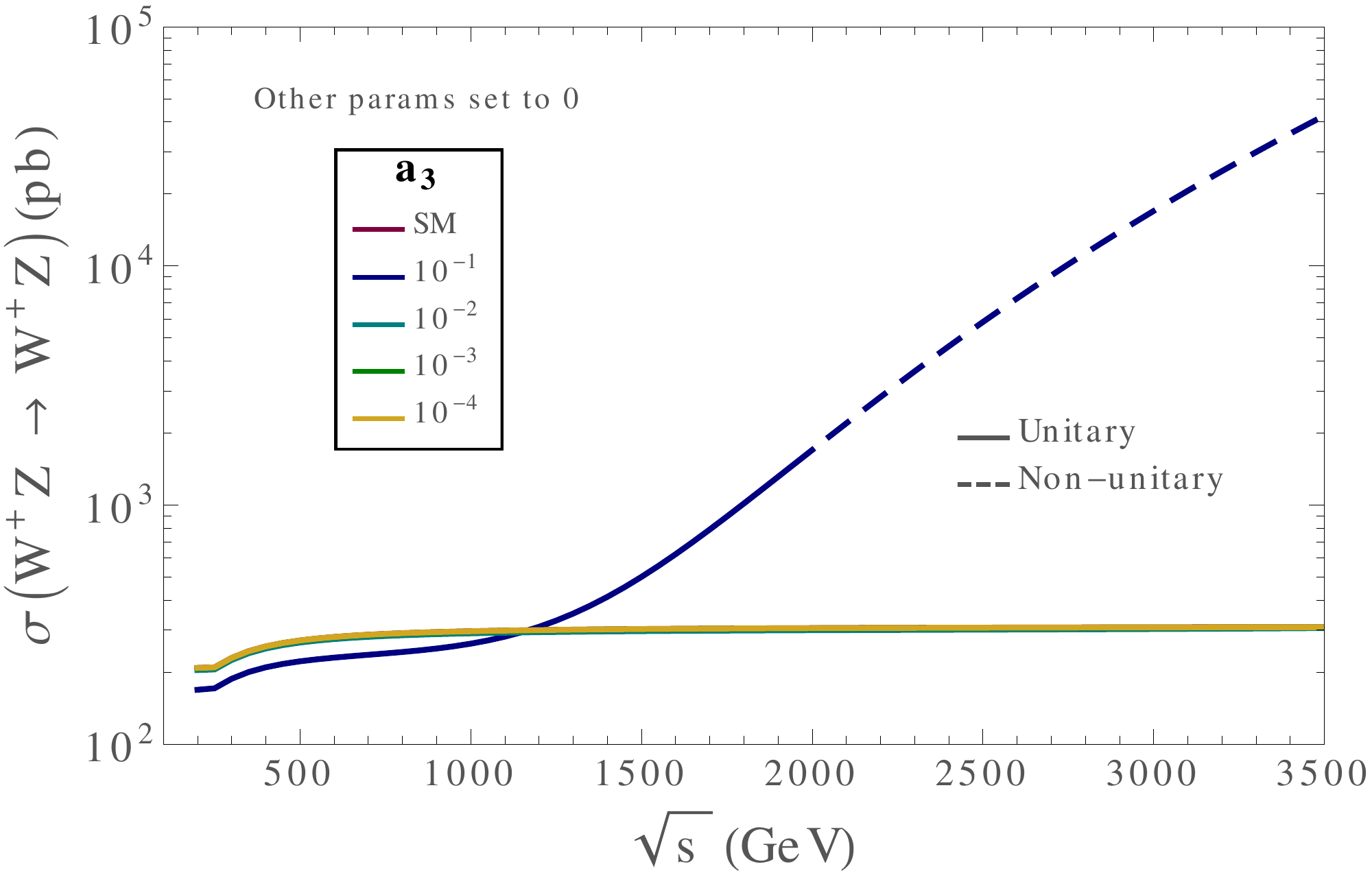}\\
\includegraphics[width=0.49\textwidth]{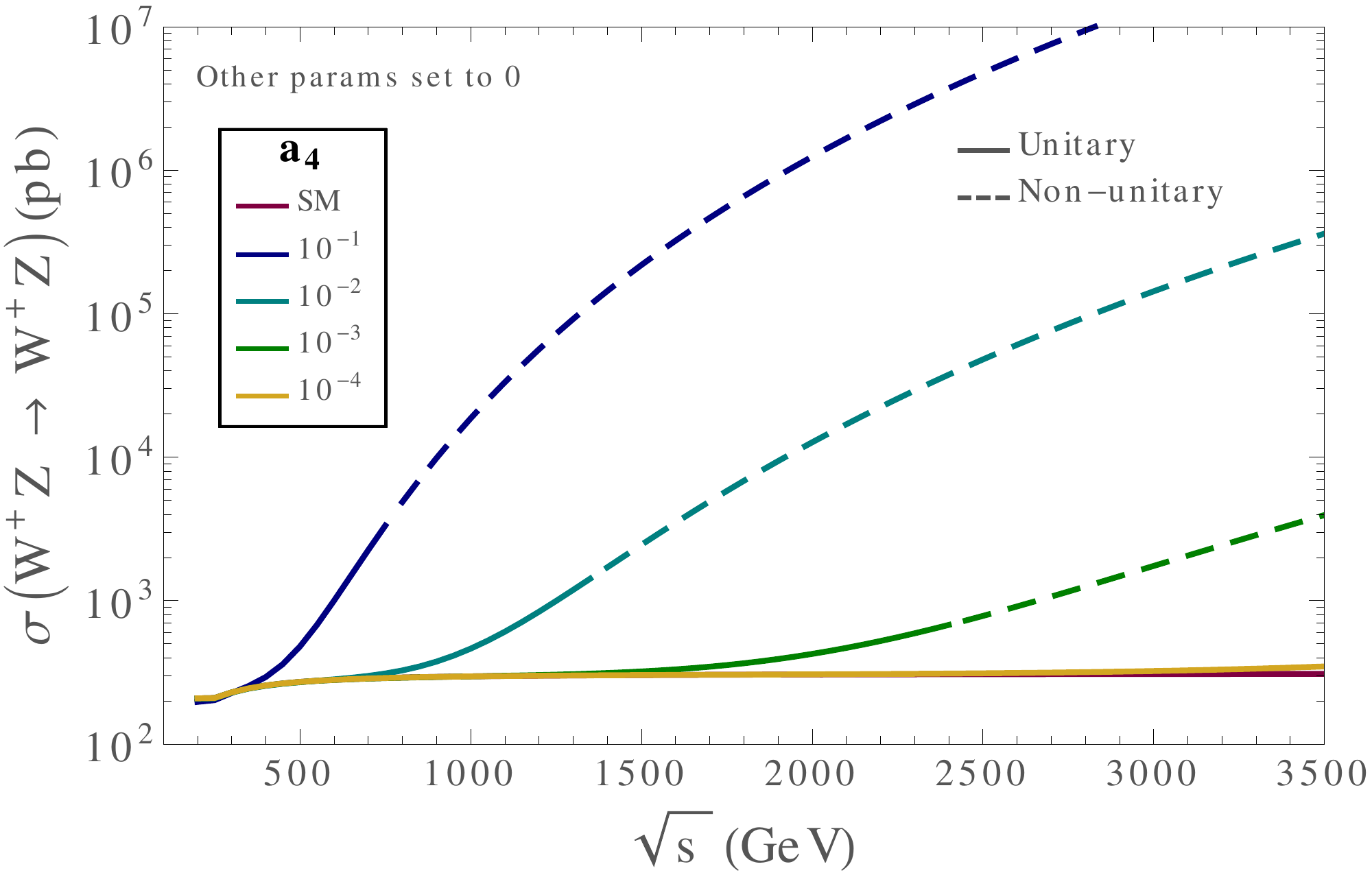}
\includegraphics[width=0.49\textwidth]{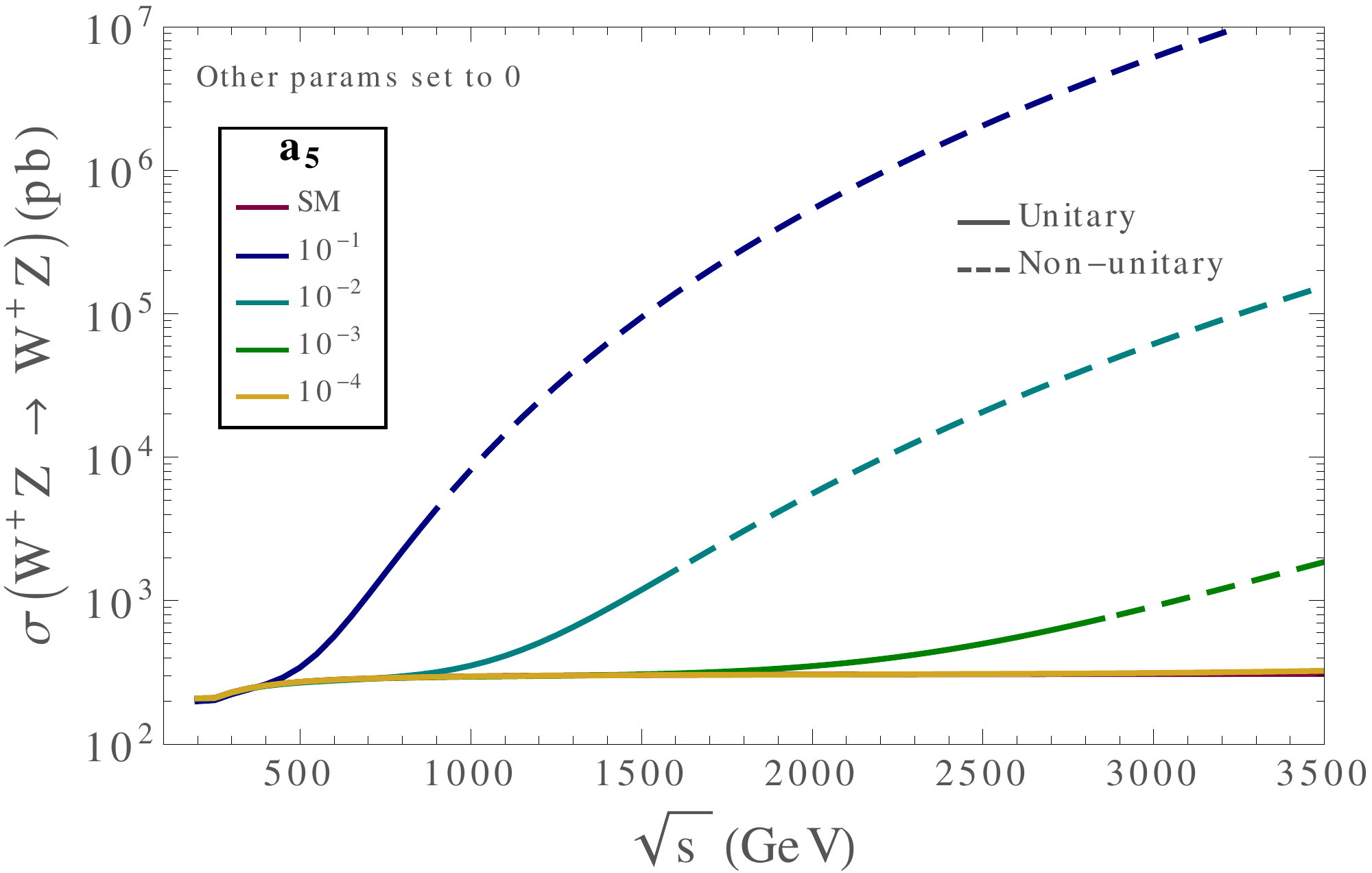}\\
\caption{Predictions of the total cross section of the process $W^+Z\to W^+Z$ as a function of the center of mass energy computed in the EChL framework for different values of one of the chiral parameters at a time. The rest are set to their SM value for a simpler comparison. From top to bottom and left to right $\Delta a\equiv a-1, a_1,a_2,a_3,a_4$ and $a_5$ are varied, respectively. Solid lines represent a unitary prediction whereas dashed lines denote unitarity violating values. Lines non visible in these plots are under the yellow line.}
\label{fig:unitviolai}
\end{center}
\end{figure}

With these considerations in mind, we now want to understand the relevance of each  EChL parameter in the violation of unitarity. Due to the fact that every of these low energy parameters has a different role in the scattering of EW gauge bosons, each of them will have a different impact on this issue. In order to characterize the violation of unitarity induced by each of the  EChL  parameters presented in Eqs. (\ref{eq.L2}) and (\ref{eq.L4}), we compute the total cross section of WZ $\to$ WZ scattering in the EChL at the tree level for different representative values of one parameter at a time, setting the rest of them to their SM value. In \figref{fig:unitviolai} we show these cross sections as a function of the center of mass energy of the process and we mark the unitarity-violating predictions with dashed lines. The value of the energy at which each cross section overcomes the unitarity limit is chosen as the lowest one at which any of the corresponding $J$ and/or helicity  partial wave crosses the unitarity bound defined in \eqref{unitlim}. In these plots it can be clearly seen that in this scattering process the parameters $a, ~a_1$ and $a_2$ (upper left, upper right and middle left panels respectively) do not play a relevant role in the violation of unitarity, since in the whole energy range that has been studied in this work there is no unitarity violation driven from these coefficients. Notice that the $b$ parameter, which controls the interaction between two EW gauge bosons and two Higgs bosons, does not appear in this scattering at tree level. When the parameter $a_3$ is considered (middle right panel), however, cross sections show a unitarity violating behavior in this same energy range. This happens only for large values of $a_3$, of the order of 0.1, for which unitarity is violated at around 2 TeV. However, this size of 0.1 is already at the border of being in conflict with the EW precision data and, therefore, a realistic choice for this $a_3$ parameter should assume a smaller value than this, leading in consequence to non-violation of unitarity in the energy range relevant for VBS at the LHC. Overall, it is clear that $a_4$ and $a_5$ are the most relevant parameters regarding the issue of the violation of unitarity in this channel. If one takes a look at the two lower panels of \figref{fig:unitviolai} it is manifest that for values of these two parameters between $0.1$ and $10^{-3}$, the violation of unitarity occurs well inside the energy range considered in this work. Actually, for values of the order of 0.1, the crossing of the unitarity limit takes place at really low values of the energy $\sqrt{s}\sim800$ GeV. Another interesting feature to pay attention to is that, in this particular channel, $a_4$ has a bigger impact in the cross section values than $a_5$. 

Based on these results, from now on we will consider only $a_4$ and $a_5$, since their contribution to the violation of unitarity is, by far, the most relevant one in this context. This is indeed intuitive, as the effective operators related to these two parameters are the only ones that remain present in ${\cal L}_4$ if    the EW gauge interactions were \textit{switched off}, i.e., in the limit $g,g' \to 0$ that corresponds to just keeping the self-interactions among the scalar modes. This is similar to the situation in chiral perturbation theory (ChPT) of low energy QCD if the electromagnetic gauge interactions were \textit{switched off} by taking the limit $e \to 0$ in the chiral Lagrangian which leads to just keeping the self-interactions among pions that provide the dominant contributions in pion-pion scattering. In summary, this means that within the EChL, if the underlying UV theory is strongly interacting, the  dominant contributions from  BSM physics to the scattering of longitudinal EW bosons are expected to be provided mainly by $a_4$ and $a_5$. Thus, and although we of course work in the framework in which the electroweak interactions are still present (i.e. we consider $g$ and $g'$ non-vanishing), relying on the assumption that the EW chiral coefficients $a_4$ and $a_5$ parameterize most relevantly the deviations from BSM physics in VBS, and specially  in terms of the violation of unitary, is well justified.

At this point, it is important to make a comment on the experimental constraints imposed on these parameters. Nowadays, several experimental studies have been devoted to set bounds on the values of these and other EFT coefficients. Here, we will discuss those concerning $a_4$ and $a_5$. For a summary on the bounds imposed upon other EChL parameters see, for instance, \cite{Delgado:2017cls}. Regarding the $a_4$ and $a_5$ constraints, there is not a unique value for them given by the LHC experimental collaborations. Although all focus their searches on VBS observables, and on the search for anomalous quartic EW gauge boson coupling signals, they differ in the interpretation of the data to extract the bounds on the EFT parameters. 

The most recent experimental searches at the LHC with $\sqrt{s}=13$ TeV aimed to constraint the parameter space of effective field theories for EWSB are explained in \cite{Sirunyan:2019der,Aad:2019xxo}. In \cite{Aad:2019xxo} a maximum total cross section of various VBS processes, and, therefore, a model independent experimental study is reported, whereas in \cite{Sirunyan:2019der} direct bounds on the linear counterparts of some EChL parameters are provided. Concerning the results in \cite{Sirunyan:2019der}, the translation to the $a_4$ and $a_5$ 95\% C.L. constraints corresponds to:
\begin{align}
 |a_4| < 6\cdot 10^{-4}\,,~~~|a_5| < 8\cdot 10^{-4}\,.
\end{align}
These are obtained without unitarizing the EChL (or, in those references the linear EFT\footnote{The relevant  parameters considered in the present work and the parameters of the linear EFT are related by $a_4=v^4/16\cdot (f_{S0}/\Lambda^4)$ and $a_5=v^4/16\cdot (f_{S1}/\Lambda^4)$}) predictions at all, through a combined study of different VBS channels and analyzing the effect of each parameter at a time. One should keep in mind that these values for the $a_4$ and $a_5$ bounds might be overestimated, since the issue of the violation of unitarity has been neglected in the corresponding study. The same happens in \cite{Sirunyan:2019ksz}, where only the WZ channel is considered in order to provide experimental bounds on $a_4$ and $a_5$. We shall not directly use these bounds as reference, since they do not take into account the issue of the violation of unitarity, but we are planning to do it in a future work.

\begin{figure}[t!]
\begin{center}
\includegraphics[width=0.5\textwidth]{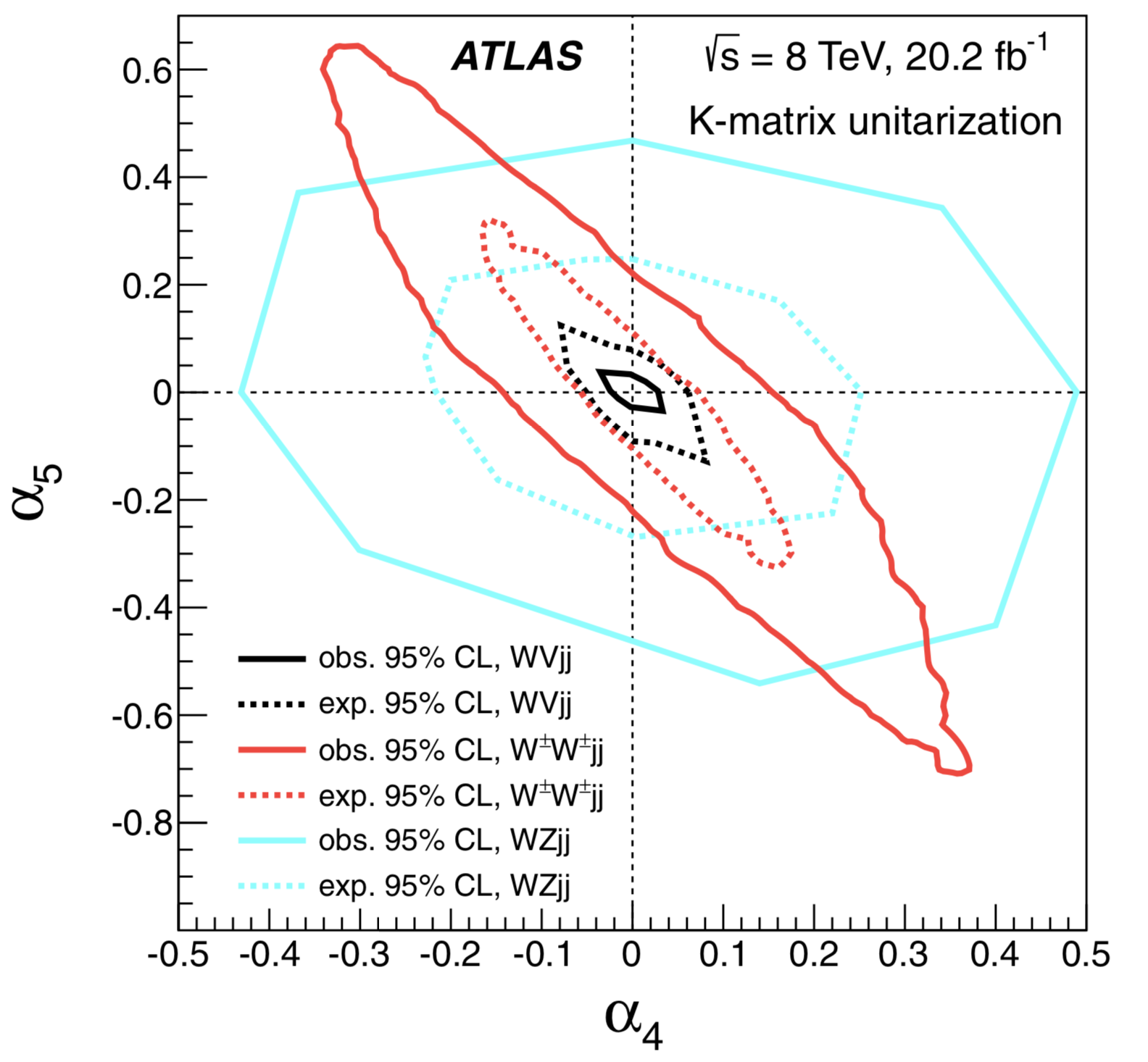}
\caption{The observed $[a_4,a_5]$ 95\% C.L. region for W${}^{\pm}$W${}^{\pm}$ final state (solid red contour), for WZ final state (solid cyan contour) and for the combined analysis (solid black contour) observed by the ATLAS collaboration interpreting the data using the K-matrix unitarization at $\sqrt{s}=8$ TeV and $L=20.2~{\rm fb}^{-1}$. The expected confidence regions are shown as well. Figure borrowed from \cite{Aaboud:2016uuk}, where the notation of  
$[\alpha_4,\alpha_5]$ is used instead of $[a_4,a_5]$.}
\label{fig:elipseATLAS}
\end{center}
\end{figure}

Another interesting bound on $a_4$ and $a_5$ is the one provided in \cite{Aaboud:2016uuk} for the LHC run at $\sqrt{s}=8$ TeV. There, a K-matrix unitarization analysis, following the procedure proposed in \cite{Alboteanu:2008my}, is performed, and the EChL $[a_4,a_5]$ parameter space is constrained, as it is shown in \figref{fig:elipseATLAS}, borrowed from \cite{Aaboud:2016uuk}. We will rely mainly upon this experimental search of \cite{Aaboud:2016uuk} as a first example in order to concrete quantitatively our final conclusions. Besides, as the overall constraints imposed in the EChL parameters in this study are of the order of $a_4\sim a_5\sim0.01$, we will use these values as reference to illustrate different VBS features without loss of generality.

In this section we have defined and studied the violation of unitarity in WZ $\to$ WZ scattering in the context of the EChL. This violation of unitarity takes place at values of the energy of ${\cal O}$(TeV) that are accesible  at the LHC for different values of the most relevant parameters, $a_4$ and $a_5$, which are the ones we will base our study on. The fact that unitarity is not fulfilled at those energies is not compatible with the correct interpretation of experimental data in order to test the EFT, since we need unitary predictions to be consistent with the underlying quantum field theory. To obtain these unitary predictions something must be done in the proper way. This is precisely to what the next section is devoted.


\section{Restoring Unitarity in WZ $\boldsymbol{\to}$ WZ scattering}
\label{Unitarity}

In the previous section we have stated that the EChL, and specially the operators governed by $a_4$ and $a_5$, lead to unitarity violation predictions for WZ $\to$ WZ scattering cross sections in the energy range accesible by the LHC. However, in order to make the EFT testable at colliders, we need to solve this problem and obtain fully unitary results for the relevant observables. To this aim, unitarization methods are addressed: prescriptions to construct unitary scattering amplitudes from the raw, non-unitary, EFT predictions. This is what we will do in this section, but,
before entering in the specific details of these unitarization methods, some general considerations have to be commented.

First of all, it is important to have in mind that relying on a particular unitarization method for the EFT implies to make some assumptions about the UV complete theory. There is therefore a trade between obtaining unitary predictions for observables and losing some of the model independence inherent to EFTs. Nevertheless, there is a caveat in this statement. When the EFT includes by construction the presence of resonant heavy states in the spectrum the various unitarization methods for VBS usualy provide comparable results, since the main features of the resonances (mass and width) are present in all cases. However, when the resonances are instead generated dynamically by the unitarization method itself (as it is the case of the Inverse Amplitude Method) this is not the case anymore and the results may vary substancially from one method to another one. Nevertheless, it is important to notice that if the unitarization methods provide amplitudes with the proper analytical structure, they can all accommodate dynamically generated resonances whose mass and width are predicted to be more or less the same independently of the employed method.  Furthermore, in the different case of non-resonant scenarios, i.e., when there are not clear emergent peaks in VBS and one searches for smooth deviations from the SM continuum,  different unitarization methods can lead to outstandingly different predictions for diverse observables. This suggests that, in order not to lose the appealing model independence of EFTs in the non-resonant case, the predictions given from the different unitarization methods available have to be contrasted, and a quantitative estimate of their differences should be provided. This inevitably introduces a theoretical uncertainty in the unitarized EFT predictions, which is precisely the one we want to quantify in this work. Therefore, we will focus in the case in which new resonant states do not manifest in the energies we are going to explore at the LHC via VBS. Besides, if present,  they would also suppose a completely different experimental setup and search strategy. 
For a recent study of these emergent resonances at the LHC via WZ scattering within the EFT approach see, for instance, \cite{Delgado:2017cls}.

Second of all, if we recall the unitarity condition given in \eqref{unitarity} that all unitarized amplitudes must fulfil, we see once again that the unitarity of a particular helicity channel does not depend just on itself but in other helicity amplitudes as well. This implies that considering only the most pathological of these amplitudes in terms of the violation of unitarity, could mean that we are neglecting important effects given the fact that the helicity system is coupled. In general, the most worrying helicity channel regarding the violation of unitarity will be the purely longitudinal one ${\rm W}_L{\rm Z}_L\to{\rm W}_L{\rm Z}_L$. This is easily understood since the longitudinal modes of the EW gauge bosons are directly related to the strongly interacting Goldstone bosons. Thus, the bigger the number of longitudinally polarized gauge bosons involved in the scattering, the lower the energy at which unitarity will be violated. In any case, the fact that the purely longitudinal helicity channel dominates at high energies and dominates the violation of unitarity depends on the particular setup that one is considering.

By studying the partial waves with the lowest values of angular momenta, $J=0,1,2$, for the 81 helicity channels independently and for different values of $a_4$ and $a_5$ and at different center of mass energies, one can disentangle the relevance of the purely longitudinal case with respect to the other helicity channels. These three lowest-order partial waves are the ones that should contain all the unitarity violating effects. This fact can be understood through the Equivalence Theorem (ET) that relates de EW gauge boson scattering amplitudes with the scalar scattering amplitudes at energies well above the EW scale. In the scattering involving just scalars in the external legs, since we have a polynomial expansion in energy up to order $s^2$ once we compute with the EChL at order $\mO(p^4)$, all partial waves with $J>2$ project to 0. Thus, the unitarity violation arising from the strongly interacting character of the interactions among scalars
 must be encoded in just the three mentioned partial waves even if we consider full gauge bosons in the external legs of our computations.

\begin{figure}[t!]
\begin{center}
\includegraphics[width=0.324\textwidth]{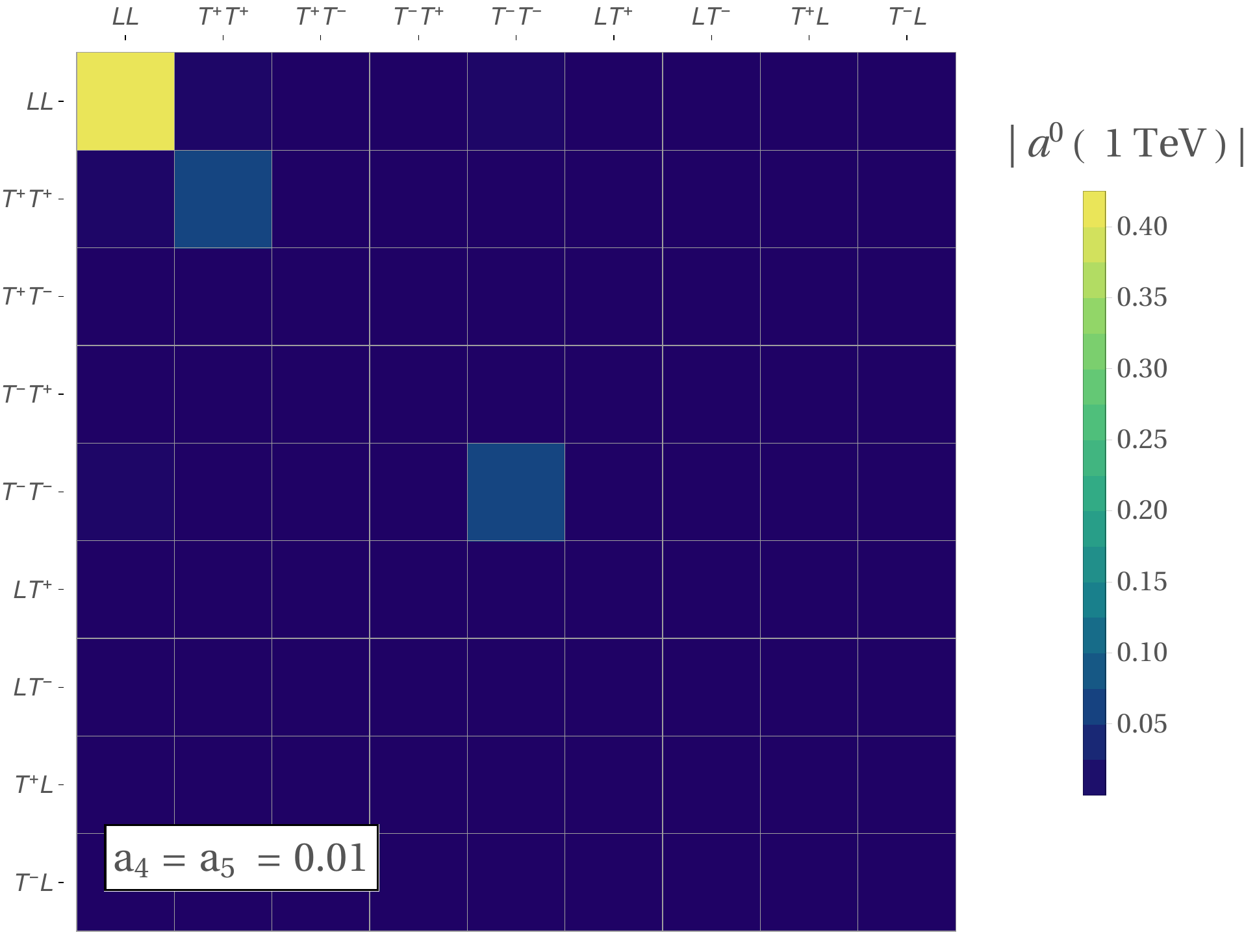}
\includegraphics[width=0.324\textwidth]{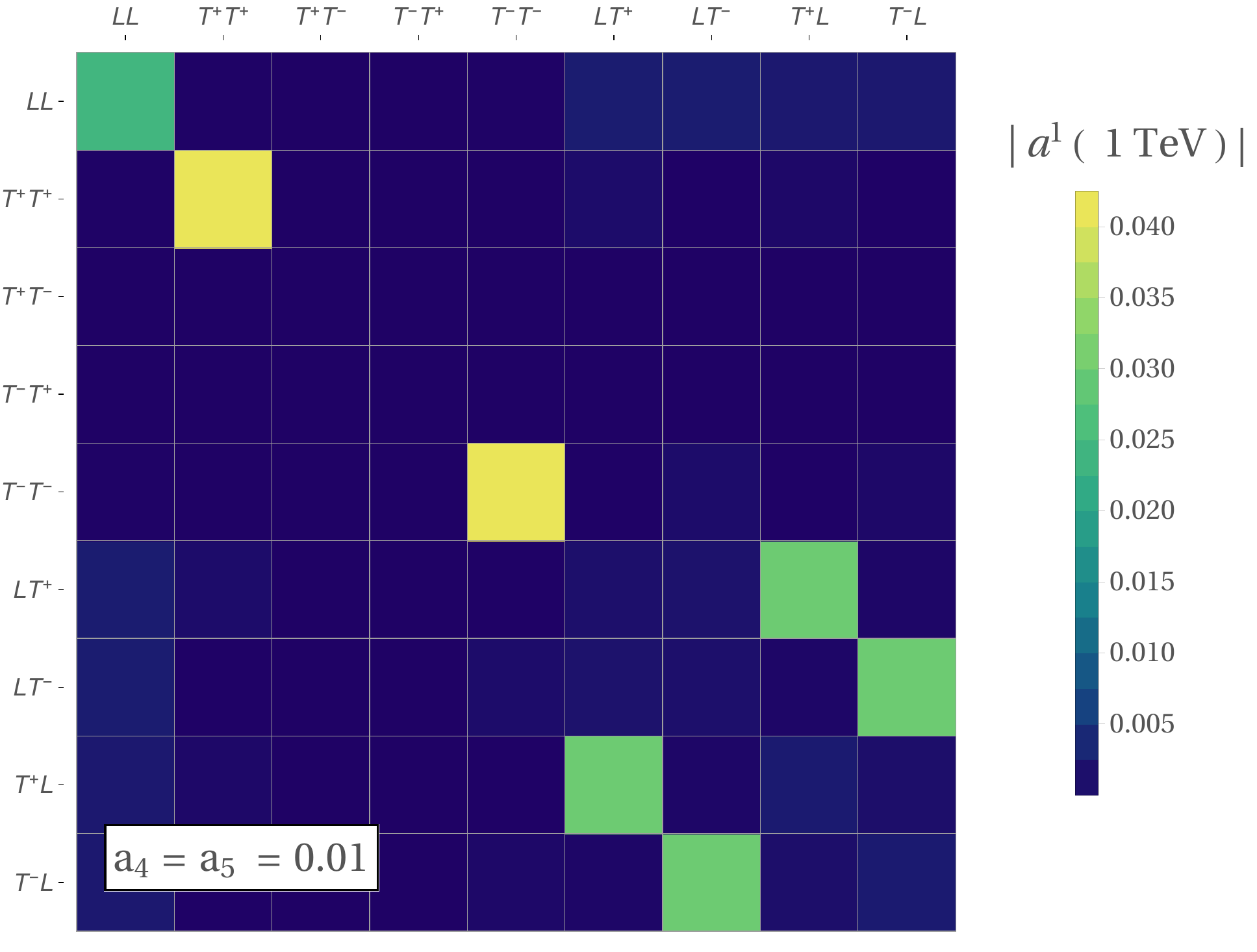}
\includegraphics[width=0.324\textwidth]{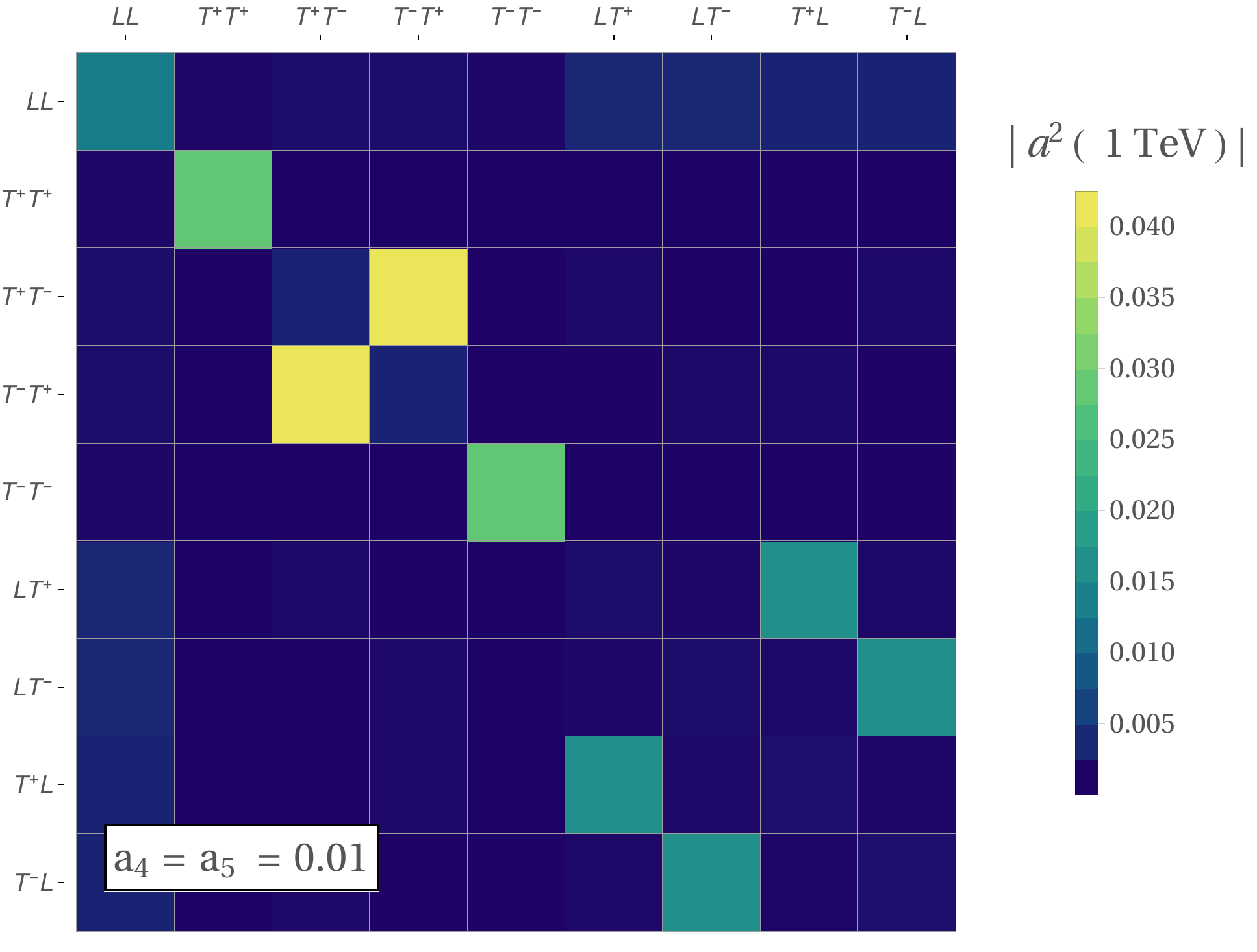}
\caption{Numerical values of the three lowest angular momentum partial waves $a^{J}(\sqrt{s})$ with $J=0$ (left), $J=1$ (middle), and $J=2$ (right), of the 81 helicity combinations of $W^+Z\to W^+Z$ scattering. Predictions are shown for a fixed center of mass energy of $\sqrt{s}=1$ TeV and for $a_4=a_5=0.01$ (with the other parameters set to their SM value) as reference. Incoming and outgoing states can be interpreted indistinctly since the results are presented in a symmetric way due to time-reversal invariance. The included labels of these 9 incoming WZ and 9 outgoing WZ states with two polarized gauge bosons, longitudinal ($L$) and/or transverse ($T^{+,-}$), are ordered and denoted here correspondingly by: $LL$, $T^+T^+$,  $T^+T^-$, $T^-T^+$ , $T^-T^-$, $LT^+$, $LT^-$, $T^+L$ and $T^-L$.}
\label{fig:teselas}
\end{center}
\end{figure}

With this in mind, we have calculated the absolute value of the three lowest-order partial waves for all the helicity channels at a certain center of mass energy and for a particular value of $a_4$ and $a_5$, in order to understand the implication of the different helicity amplitudes in the total cross section and in the unitarization process. In \figref{fig:teselas} we present an example of this for the reference values of $a_4=a_5=0.01$ and for a representative center of mass energy of 1 TeV. Looking at this figure, one can observe various interesting features. The first one is that, in general terms, the $J=0$ partial wave (left panel) is around one order of magnitude bigger than the other two, $J=1,2$ (middle and right panels, respectively), as it is already well known in the literature. The second one is that only for that same value of the angular momentum, $J=0$, the purely longitudinal scattering (displayed in the (1,1) entry of these ``matrices'', where incoming and outgoing states can be interpreted indistinctly since the results are presented in a symmetric way due to time-reversal invariance) dominates, being it a factor 5 larger than the next contributing helicity channel and thus becoming practically the only relevant amplitude to take into account. In the other two cases, $J=1,2$, the $LL\to LL$ case is no longer dominating the picture and other helicity channels become important. In particular we see in this figure that $T^+T^+\to T^+T^+$ and $T^-T^-\to T^-T^-$ play a relevant role in $J=1$ and $T^+T^-\to T^-T^+$ and $T^-T^+\to T^+T^-$ do it in $J=2$. This points towards the fact that, in some setups and for determined values of the relevant chiral parameters, neglecting the unitarity-violating effects of channels other than the purely longitudinal one could lead to incomplete predictions. This is the reason why we will consider the whole coupled system of the 81 helicity amplitudes when applying the mentioned unitarization methods.

The third comment one has to make regarding unitarization methods is somehow obvious, but important: all unitarization schemes have to provide similar predictions in the low energy region, i.e., above but not far from the WZ threshold production. This is a well known feature in the context of ChPT where the scattering amplitudes from the chiral Lagrangian,  unitarized with the various methods, do recover the ChPT prediction at low energies, in agreement with the well known low-energy theorems. 

Having stated all these considerations, we proceed to briefly explain the unitarization methods that we are going to consider in the present work. We have selected them based on the fact that they are the most used ones nowadays in the literature. Since these methods are the ones that are currently being used to interpret the experimental data in order to obtain information about the EFT, we find pertinent to contrast their predictions. They can be classified in two categories: 1) the ones that directly suppress {\it by hand} the pathological energy behavior of the amplitudes with energy (that we call here, as it is usual in the literature, Cut off, Form Factor and Kink), and 2) the ones that unitarize 
the first three partial waves from which then the total unitary amplitude is reconstructed (K-matrix and Inverse Amplitude Method (IAM)). Furthermore, they differ in their physical implications and motivation, and in their analytical properties, that we will discuss in the next paragraphs. Despite these differences, and the fact that some of them could be more physically justified than others, there is in principle no prior to choose a particular method with respect to the others.  

We now list the five unitarization prescriptions considered in these work with a brief explanation of each of them (for some illustrative reviews on different unitarization methods in the context of VBS see, for instance, \cite{Kilian:2014zja, Rauch:2016pai,DelgadoLopez:2017ugq}) .

\begin{itemize}
\item{\textbf{Cut off:}
The Cut off is not a unitarization method per se but a way to obtain unitary amplitudes by just discarding those predictions given for energy values above the unitarity violation scale $\Lambda$, defined in the previous section as the lowest value of $\sqrt{s}$ at which any partial wave crosses the unitarity bound stated in \eqref{unitlim}. This would mean to reject the predictions  of the cross sections marked with dashed lines in \figref{fig:unitviolai}, sticking only to those that respect the unitarity condition (i.e., solid lines in these figures).
}
\item{\textbf{Form Factor (FF):}
In this case, instead of obviating part of the results computed from the raw EFT, what is done is to suppress the pathological behavior of the amplitudes with energy above the scale at which each of them violate unitarity. To that purpose, a smooth, continuous function of the form:
\begin{align}
f_i^{FF}=(1+s/\Lambda_i^2)^{-\xi_i}
\end{align}
is employed. Here $s$ is the center of mass energy squared, $\Lambda_i$ is the specific value of $\sqrt{s}$ at which the helicity channel $i$ violates unitarity according to \eqref{unitlim} and $\xi_i$ is the minimum exponent that is sufficient  to fix the pathological behavior of the corresponding $i^{th}$ helicity amplitude with energy. Thus, every non-unitary helicity amplitude will be unitarized in the following manner:
\begin{align}
\hat{A}_{\lambda_1,\lambda_2,\lambda_3,\lambda_4}=A_{\lambda_1,\lambda_2,\lambda_3,\lambda_4}\cdot(1+s/\Lambda_{\lambda_1,\lambda_2,\lambda_3,\lambda_4}^2)^{-\xi_{\lambda_1,\lambda_2,\lambda_3,\lambda_4}}\,,\label{FF}
\end{align}
with $\hat{A}$ being the unitary amplitude and $A$ the non-unitary EFT prediction. With all these unitarized amplitudes, then, one would be able to recover a unitary unpolarized, total cross section. In the present case, and for the values of the chiral parameters that are going to be probed in this work, the scales at which unitarity is violated for all helicity channels are above the maximum center of mass energy considered, except in the purely longitudinal case. We have checked that including the Form Factor suppression given in \eqref{FF} for all helicity channels (notice that not only the scale is different in each channel, but also the exponent since they depend differently with energy) is equivalent to do it just in the $LL\to LL$ one for the energies and parameters we are considering, so, for simplicity, from now one we will apply \eqref{FF} to the scattering of longitudinally polarized gauge bosons leaving the rest unchanged. In this way, our prescription to apply the Form Factor unitarization method can be summarized as:
\begin{align}
\hat{A}_{LLLL}=A_{LLLL}\cdot(1+s/\Lambda_{LLLL}^2)^{-2}\,,\label{FFLLLL}
\end{align}
recalling that any other helicity amplitude is left unaffected. The exponent has been set to $\xi_{LLLL}=2$ since it is the minimum value necessary to repair the anomalous growth with energy of the $LL\to LL$ amplitude. The scale $\Lambda_{LLLL}$ has been computed with the VBFNLO utility to calculate Form 
Factors~\cite{Arnold:2008rz,Arnold:2011wj,Baglio:2014uba}.
}
\item{\textbf{Kink:}
The so called Kink unitarization method is very similar to the Form Factor. Conceptually, it is the same, and the only difference present between both prescriptions is that the suppression in the Kink method is not performed smoothly, but with a step function:
\begin{align}
f_i^{Kink}=\left\{\begin{array}{l}1~~~~~~~~~~~~~s\leq\Lambda_i^2 \\(s/\Lambda_i^2)^{-\xi_i} ~~ s>\Lambda_i^2\end{array}\right.\,.
\end{align}
Except for this fact, the rest of the discussion regarding the Form Factor is equally valid for the Kink, so, in this case, we will also apply the method only to the $LL\to LL$ amplitude with an exponent of $\xi_{LLLL}=2$.
}
\item{\textbf{K-matrix:}
The K-matrix unitarization method has been extensively studied and implemented in the context of ChPT  in QCD. This method is a prescription applied to the partial wave amplitudes and basically projects the non-unitary ones into the Argand circle through a stereographic projection. This means that it takes a real, non unitary partial wave amplitude to which an imaginary part is added {\it ad hoc} such that the unitarity limit is saturated. For each helicity partial wave amplitude, this is achieved by using the following simple formula:
\begin{align}
\hat{a}^{J;{\rm K-matrix}}_{\lambda_1\lambda_2\lambda_3\lambda_4}=\dfrac{a^J_{\lambda_1\lambda_2\lambda_3\lambda_4}}{1-i\,a^J_{\lambda_1\lambda_2\lambda_3\lambda_4}}\,.
\end{align}
However, as we have already commented throughout the text, the unitarity condition implies that the whole coupled system of helicities has to be taken into account in our unitarization procedures. Thus, we solve this coupled system in terms of matrices, for which we construct a 9$\times$9 matrix, whose entries correspond to the 81 possible helicity amplitudes of the elastic WZ scattering we are studying, and we unitarize it using the K-matrix method. This way we have:
\begin{align}
\hat{\alpha}^{J;{\rm K-matrix}}=\alpha^J\cdot[1-i\,\alpha^J]^{-1}\,,
\end{align}
being $\alpha$ the 9$\times$9 matrix containing the whole system of helicity partial wave amplitudes. Now, what we need is to reconstruct, from these unitary partial waves, the complete scattering amplitude. To this aim, we substitute from the initial, non-unitary amplitude, the unitarity violating partial waves by their unitarized versions. As we have already explained in the text, these partial waves are those that correspond to $J=0,1,2$, so, what we do is to subtract these three partial waves from the total amplitude to then add the same partial waves after the K-matrix unitarization has been performed:
\begin{align}
\hat{A}_{\lambda_1\lambda_2\lambda_3\lambda_4}(s,\cos\theta)=&\,A_{\lambda_1\lambda_2\lambda_3\lambda_4}(s,\cos\theta)-16\pi\sum_{J=0}^2 (2J+1)\,d^J_{\lambda,\lambda'}(\cos\theta)\,a^J_{\lambda_1\lambda_2\lambda_3\lambda_4}(s)\,\nn\\
&+16\pi\sum_{J=0}^2 (2J+1)\,d^J_{\lambda,\lambda'}(\cos\theta)\,\hat{\alpha}^{J;{\rm K-matrix}}_{[\lambda_1\lambda_2\lambda_3\lambda_4]}(s)\,.\label{amprec}
\end{align}
Here we denote as $\hat{\alpha}^{J;{\rm K-matrix}}_{[\lambda_1\lambda_2\lambda_3\lambda_4]}(s)$ (in the rest of the formulas it is implicit that all partial waves depend solely on $s$) the element of the 9$\times$9 matrix that corresponds to the $\lambda_1\lambda_2\lambda_3\lambda_4$ polarization state.  In this way, we obtain a unitary amplitude in which we maintain all the fundamental properties introduced by all the  partial wave amplitudes, including those with higher $J>2$ that, since are not involved in the violation of unitarity, remain unaffected. The numerical computations in this K-matrix case and the next one, IAM,  have been performed with a private mathematica code developed  by us.  
}
\item{\textbf{Inverse Amplitude Method (IAM):}
The Inverse Amplitude Method is, probably, the most profoundly studied unitarization prescription considered in this work. It is very well known in the context of ChPT for pion-pion scattering, and its accuracy has been proved in various scenarios, like, for instance, in the prediction of the $\rho$ meson as an emergent resonance in these scattering processes}. It is based on the application of dispersion relations (bidirectional mathematical prescriptions allowing to relate the real and imaginary parts of complex functions) to the inverse of the partial wave amplitudes computed in the EFT framework. This unitarization procedure can be actually understood as the result of the first Pad\'e approximant derived from the chiral expansion series provided by ChPT. In practice, this method implements an approximate re-summation of loops with bubbles in the s-channel of the given scattering process. Therefore in the present context of the EChL it accounts for re-scattering effects in the scattering of the two EW gauge bosons, i.e., WZ in our chosen example,  which are not taken into account with the other unitarization methods. Notice that this makes sense in the context of a strongly interacting theory since these re-scattering contributions are not suppressed as in weakly interacting systems. 

In summary, if one starts with the typical result for a given partial wave amplitude from the chiral Lagrangian, given by the sum of the two contributions in the chiral expansion, one of order ${\cal O}(p^2)$ and the other one of order ${\cal O}(p^4)$ , the corresponding prediction of the IAM  leads to the following unitarized helicity partial wave amplitudes:
\begin{align}
\hat{a}^{J;{\rm IAM}}_{\lambda_1\lambda_2\lambda_3\lambda_4}=\dfrac{(a^{{(2)\,J}}_{\lambda_1\lambda_2\lambda_3\lambda_4})^2}
{a^{(2)\,J}_{\lambda_1\lambda_2\lambda_3\lambda_4}-a^{(4)\,J}_{\lambda_1\lambda_2\lambda_3\lambda_4}}\,,
\end{align}
where $a^{(2)}$ is the contribution to the partial wave amplitude computed with the operators from the $\mL_2$ Lagrangian (\eqref{eq.L2}) at the tree level, which is of order ${\cal O}(p^2)$ and $a^{(4)}$ is the contribution to the partial wave amplitude computed with the operators from the $\mL_4$ (\eqref{eq.L4}) at the tree level, plus the contribution computed with the operators from the  $\mL_2$ Lagrangian at one loop level, which are both of ${\cal O}(p^4)$. In the present work, since the computation of the complete one loop level amplitudes that enter in $a^{(4)}$ has not been performed yet due to the difficulty of the task, we will evaluate this here in an approximate way. Following the usual features in ChPT, we take 
the imaginary part of this contribution to be $\left|a^{(2)}\right|^2$ so that the unitarity condition is fulfilled perturbatively, and neglect the real contribution of the loops which are expected to provide a very small contribution, not being relevant for the present computation. 

Once again we encounter ourselves in the scenario in which we have a prescription to unitarize each helicity amplitude independently. However, we want to take the whole coupled system of helicities in full generality, as explained above. We construct once more the 9$\times$9 matrix $\alpha$, this time splitting it into its $\mO(p^2)$ and $\mO(p^4)$ contributions, that contains the 81 helicity amplitudes, and we unitarize it using the IAM in the following matricial manner:
\begin{align}
\hat{\alpha}^{J;{\rm IAM}}=\alpha^{(2)\,J}\cdot[\alpha^{(2)\,J}-\alpha^{(4)\,J}]^{-1}\cdot\alpha^{(2)\,J}\,.\label{matrixIAM}
\end{align}
At this point, to obtain a fully unitary amplitude, we use the same trick as in the K-matrix case, i.e., we replace the unitarity violating partial waves of the total amplitude by their IAM-unitarized version, following \eqref{amprec} with the only change of K-matrix$\to$IAM. It is pertinent to make now some comments regarding important differences between the IAM unitarization method and the rest we have considered. The IAM does not provide just unitary predictions, but also succeeds to get partial wave amplitudes with the appropriate analytical structure (for more details on this, see, for instance, \cite{DelgadoLopez:2017ugq}). This implies that it is the only method, amongst the ones studied in this work, that can accommodate dynamically generated resonances, since these appear as complex poles in the second Riemann sheet of the partial wave with the corresponding $J$ quantum number. This is in contrast to the unitarized partial waves with the K-matrix method that do not have such poles. These resonances are characteristic of strongly interacting theories, and appear naturally at high energies, such as in the case of low-energy QCD. Furthermore, it is worth commenting that, according to \cite{Delgado:2015kxa}, similar results as those obtained with the IAM regarding the appearance of dynamical resonances  are also provided by other alternative unitarization methods that lead to the proper analytical structure. Example of such methods are the N/D or the improved K-matrix, which for shortness, we have decided not to include here.    
Nevertheless, for the forthcoming study at the LHC, as we have already said, we are interested in studying the non-resonant case of the unitarized theory, so the differences among the various unitarization methods will come in terms of smooth deviations from the SM continuum {\it via} WZ scattering rather than from the appearance of peaks due to the emergence of resonances. It is important, though, to keep in mind that the IAM has some peculiarities regarding its structure and physical motivation, that differentiates it from the others.
\end{itemize}

\begin{figure}[t!]
\begin{center}
\includegraphics[width=0.49\textwidth]{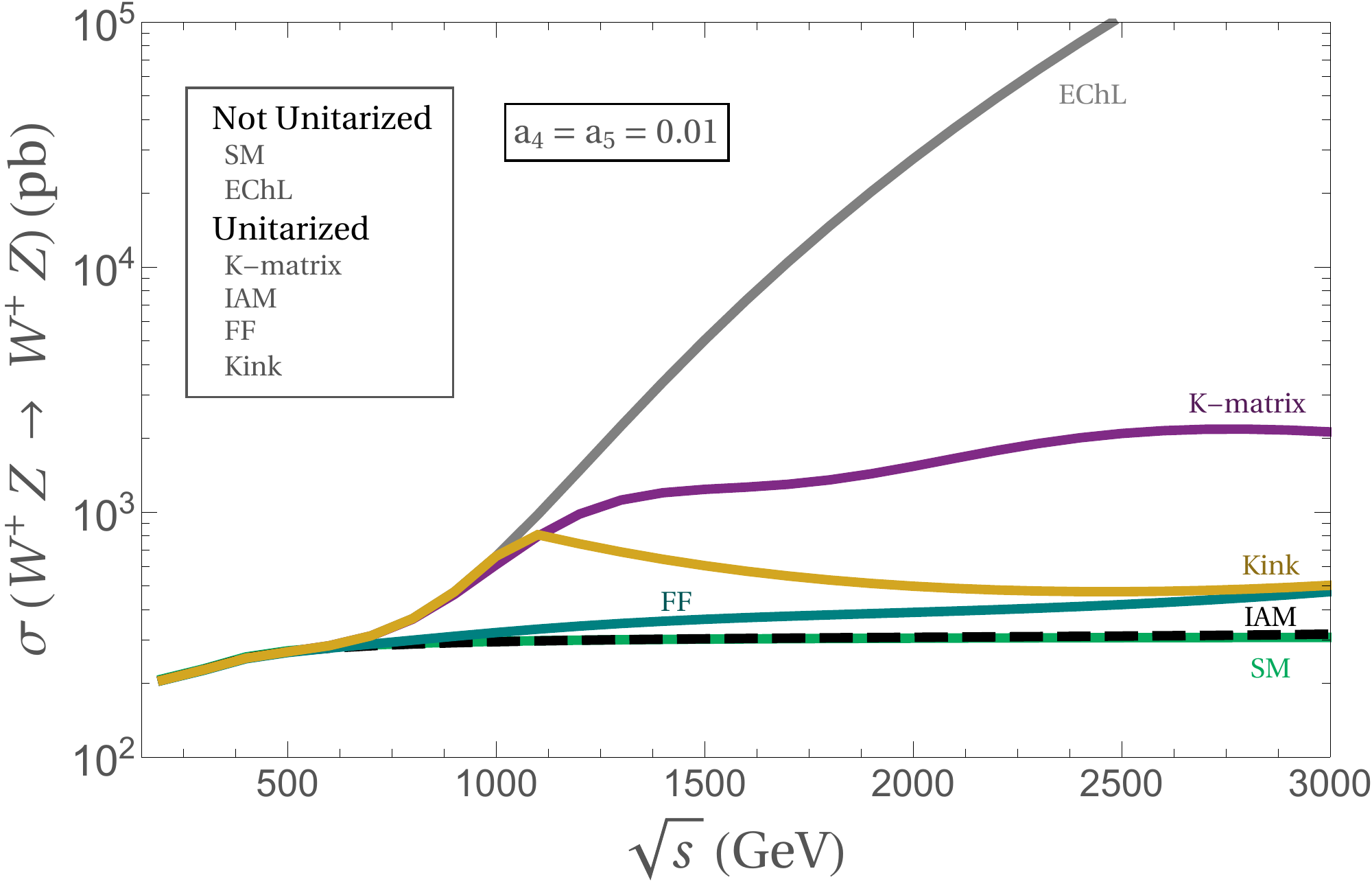}
\includegraphics[width=0.49\textwidth]{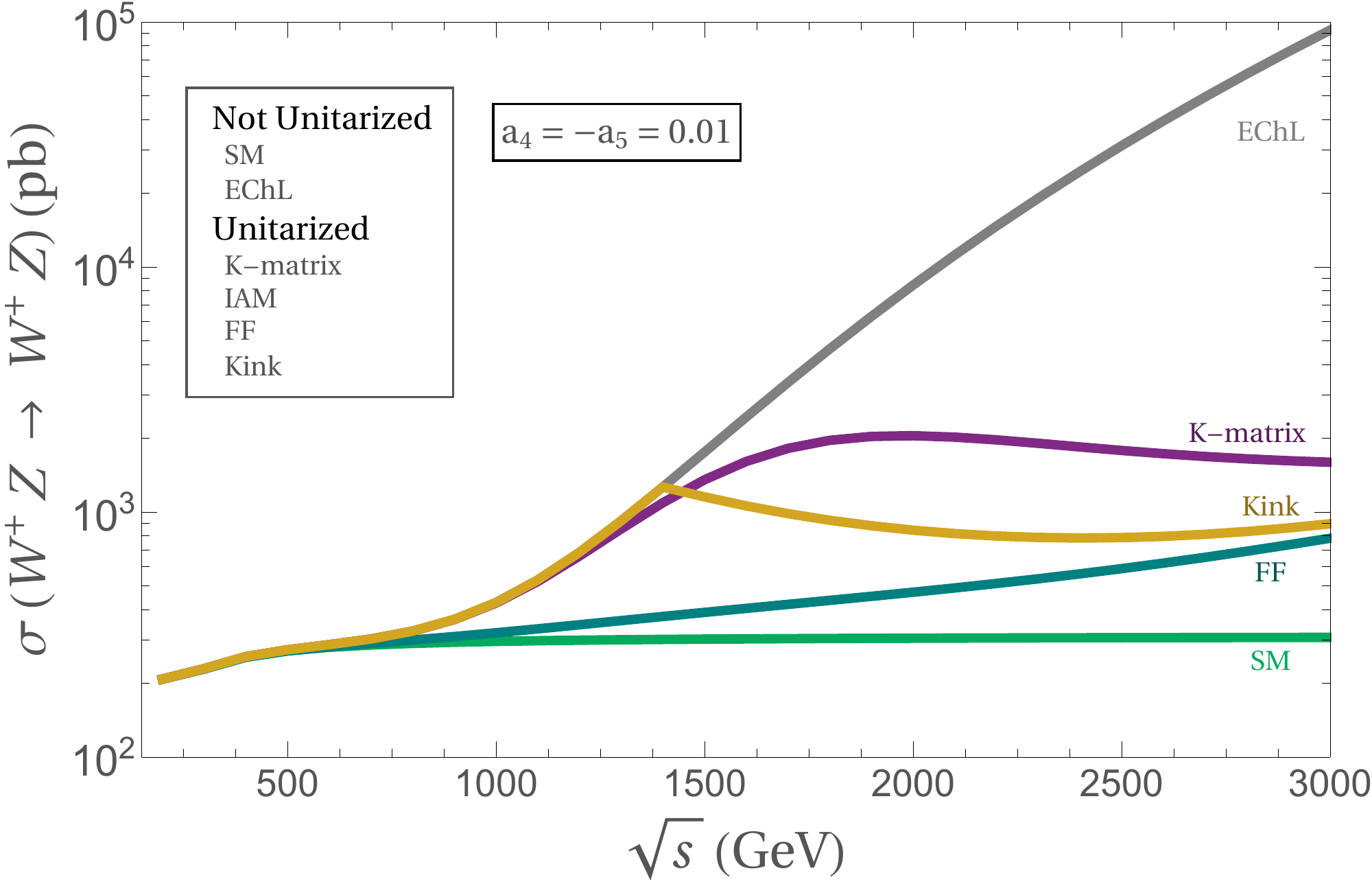}
\caption{Predictions of the total cross section of the process $W^+Z\to W^+Z$ as a function of the center of mass energy for the different unitarization procedures explained in the text: K matrix (purple), Kink (yellow), Form Factor (FF, blue) and IAM(dashed black). Non-unitarized EChL and SM are also displayed. Two benchmark $a_4,a_5$ values are displayed: $a_4=a_5=0.01$ (left) and $a_4=-a_5=0.01$ (right). In all plots $a=1$ (or, equivalently, $\Delta a=0)$. }
\label{fig:methodssubprocess}
\end{center}
\end{figure}

We have already discussed briefly each of the unitarization procedures that we consider in this work and the specific way in which we implement them. Now, what we need is to study the different predictions they provide in regard of VBS observables. To that purpose, we apply each of them, in the above explained manner, to the WZ$\to$WZ scattering amplitudes for different values of $a_4$ and $a_5$. The final results of this computation can be seen in \figref{fig:methodssubprocess}, where we present the total, unpolarized cross section of the WZ$\to$WZ scattering as a function of the center of mass energy for the different unitarization methods used. We also display the SM prediction and the non-unitarized EChL prediction for comparison. We consider, moreover, two scenarios for the values of $a_4$ and $a_5$. We set their absolute values to 0.01, as reference, and analyze two cases: the one in which both have the same sign $a_4=a_5=0.01$ (left) and the one in which they have opposite sign $a_4=-a_5=0.01$ (right).

 A great number of interesting facts can be extracted from these plots. Firstly, and most clearly, one can see that different unitarization methods lead, indeed, to very different predictions for this observable. These predictions differ, in general, from those of the raw EChL and the SM, as well. Therefore, one can expect that these differences might be very well seen experimentally. Secondly, another interesting issue arises from the comparison of both panels in the figure. It appears plainly that the value of the EChL prediction and of the 
 K-matrix prediction are, in general, smaller in the case in which both parameters,  $a_4$ and $a_5$, have opposite sign. On the contrary, the Kink and the Form Factor provide larger results in this same case. This means that for the non-unitarized prediction and for the K-matrix, the regions of the parameter space in which $a_4$ and $a_5$ have the same sign will be more constrained, whereas for the Kink and the Form Factor the opposite-sided regions will be the most constrained ones. We have checked that the predictions for the scenario in which both parameters are negative gives the same results as the one in which they are both positive, and that in the case in which they have opposite sign the same result is obtained when either of the parameters is positive/negative. Thirdly, a comment has to be made regarding the Cut off procedure. The unitarity violation scale is not explicitly shown in these plots, but it can be inferred from the position of the ``knee'' in the Kink prediction. As it is clear, discarding the values of the cross section above this scale will imply to lose a lot of sensitivity, and will of course correspond to a very different prediction with respect to the other studied cases. 
 
Regarding the IAM, we can clearly see that for the particular choice of parameters in the left panel of \figref{fig:methodssubprocess} its prediction lies very close to the SM one. In this case the IAM does not provide an emergent resonance in WZ scattering, since for these particular values of the EChL parameters there are not poles in the reconstructed total amplitude from the IAM partial waves and, as a consequence, the outcome provided by the IAM when applied to the LHC context will not show any departure from the SM continuum. In contrast, for the particular choice of parameters in the right panel, there is indeed an emergent resonance below 1 TeV,  which we have decided not to include in this plot in the right panel since it is most probably already excluded by the present searches at the LHC.

\begin{figure}[t!]
\begin{center}
\includegraphics[width=0.49\textwidth]{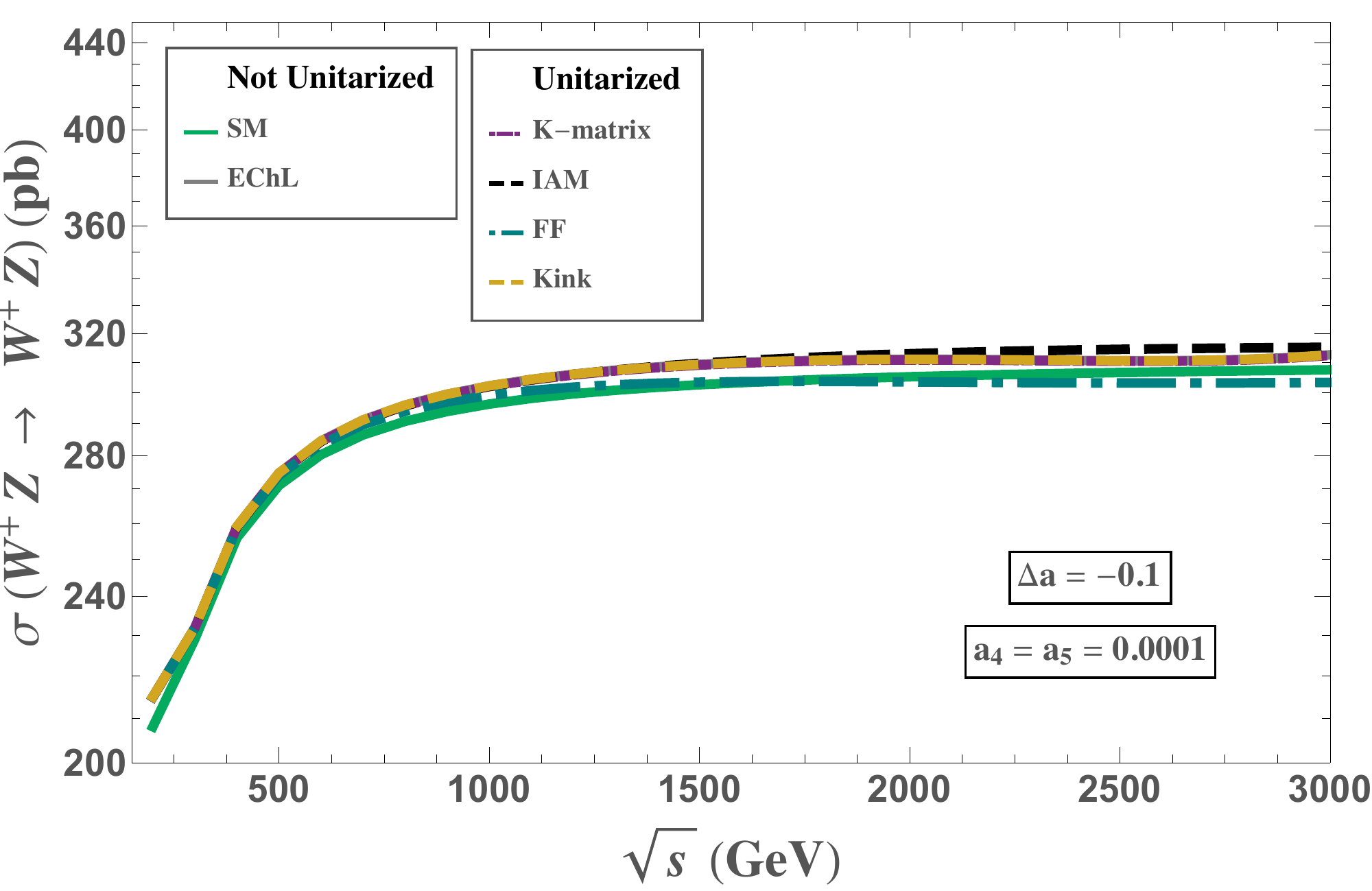}
\includegraphics[width=0.49\textwidth]{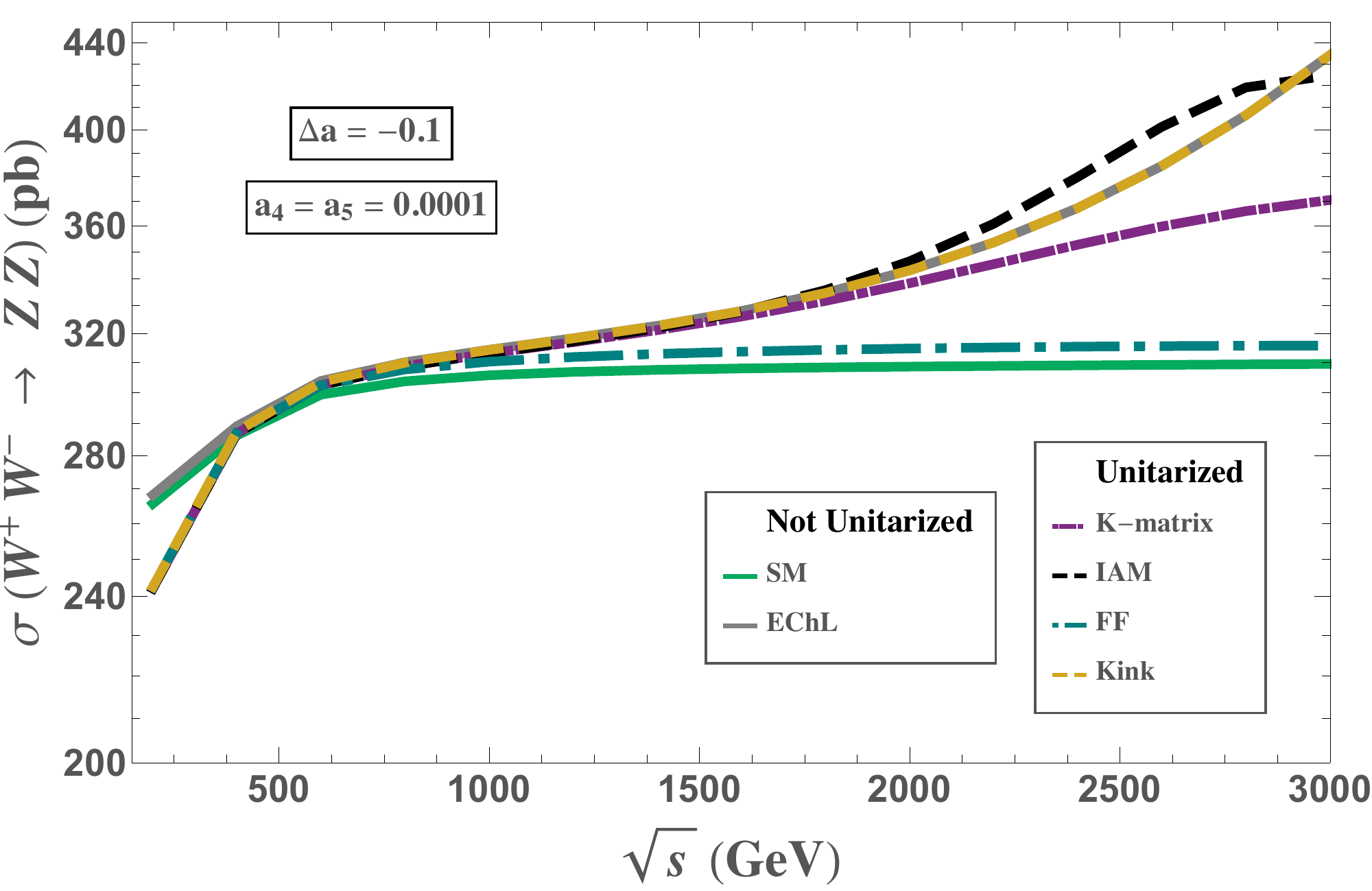}\\
\includegraphics[width=0.49\textwidth]{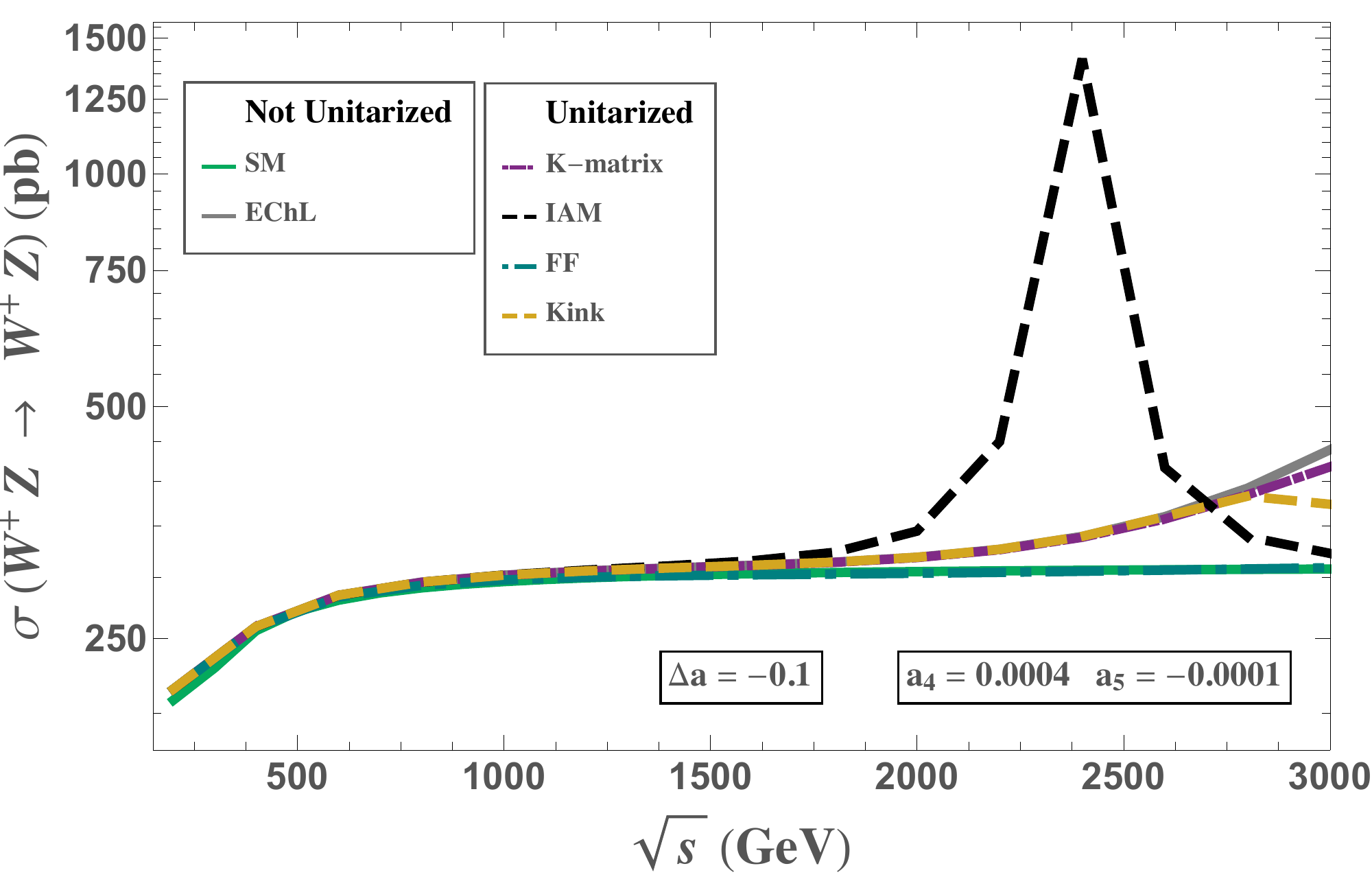}
\includegraphics[width=0.49\textwidth]{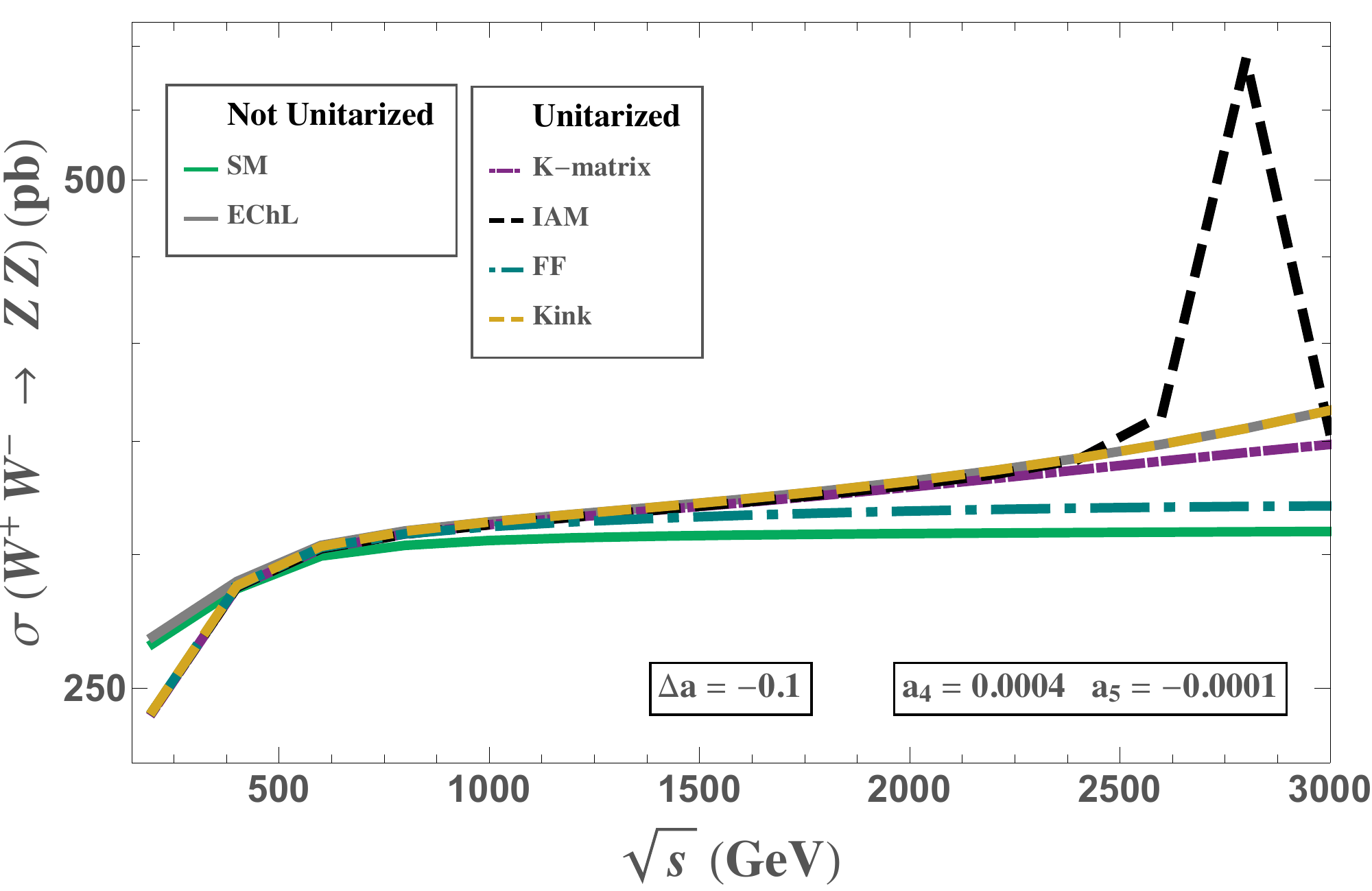}
\caption{Predictions of the total cross section of the process W${}^+$Z$\to$ W${}^+$Z (left panels) as a function of the center of mass energy for the different unitarization procedures explained in the text: K-matrix (purple), Kink (yellow), Form Factor (FF, blue) and IAM(dashed black). Non-unitarized EChL (gray) and SM (green) are also displayed. Two benchmark $a_4,a_5$ values are displayed: $a_4=a_5=0.0001$ (upper) and $a_4=0.0004,~ a_5=-0.0001$ (lower). For comparison, we include the plots corresponding to our predictions for same choice of the parameters but for the channel WW $\to$ ZZ (right panels). In all plots, $a=0.9$ (or, equivalently, $\Delta a=-0.1$.)}
\label{fig:comparisonwithwwzz}
\end{center}
\end{figure}

When other particular values of the EChL parameters are chosen,  different patterns in the predictions of the VBS cross sections from the various unitarization methods can appear. In general, the choice of smaller values of $|a_4|$ and $|a_5|$ than those in  \figref{fig:methodssubprocess} typically lead, in the non-resonant case, to closer predictions for the various unitarization methods in the studied energy range, and also closer to the SM prediction. This can be clearly seen in the upper left pannel in \figref{fig:comparisonwithwwzz} where the parameters have been set to $a_4=a_5=0.0001$ and $a=0.9$ (or equivalently, $\Delta a= a-1=-0.1$). For this particular choice,  a scalar resonance emerges close to 3 TeV in the IAM unitarized predictions, which does not manifest in the channel of our interest here $WZ \to WZ$ but in the $WW \to ZZ$ channel. This can be seen clearly in the plot of the upper right panel   in   \figref{fig:comparisonwithwwzz}, which we have included for comparison. In this case, studying this alternative VBS channel $WW \to ZZ$ at the LHC seems more appropriate in order to  analyze the distortions with respect to the SM predictions due to BSM physics represented by this particular choice of parameters.

The other example included in \figref{fig:comparisonwithwwzz}, where the parameters are set to $a_4=0.0004$, $a_5=-0.0001$ and again $a=0.9$, displays the emergence of a vector resonance in the IAM prediction for $WZ \to WZ$ (lower left panel) close to 2500 GeV, and a scalar resonance close to 2800 GeV in the IAM prediction for $WW \to ZZ$ (lower right panel). This resonant behavior is only found in the predictions with the IAM but not in the predictions with the other unitarization methods.   

In summary, regarding the IAM, the appearance of dynamically generated resonances in the energy range of a few TeV occurs indeed for a continuum set of $a_4$ and $a_5$ values of the order of ${\cal O}(10^{-3}-10^{-4})$ and its properties, mass and width, also depend on the other relevant parameters, particularly on $a$. These features of the IAM have been studied extensively in the literature and are not the main focus of the present paper which, as we have said, is mainly devoted to the non-resonant case. Thus, for the rest of this work we will focus on the other unitarization methods which will produce instead smooth distortions from the SM continuum.

Finally, a significant point has to be made concerning the K-matrix cross sections, since they are the ones we will use in the next section as a link to the experimental results. We have compared our estimates, obtained with the K-matrix procedure explained in the pages above, with the ones provided by the Wizard group~\cite{Kilian:2014zja,Alboteanu:2008my}. In the given references, the authors construct unitarized four point functions that can be introduced in a Monte Carlo event generator. Their prescription is based in the T-matrix unitarization method, that they implement in a similar way than us: replacing the unitarity violating partial wave amplitudes of the total amplitude by their T-matrix unitarized version\footnote{In the present case, their T-matrix unitarization is equivalent to our K-matrix unitarization, as we have explicitly checked.}. This prescription is used, actually, by the ATLAS collaboration in order to constrain the EChL parameter space~\cite{Aaboud:2016uuk}. Nevertheless, their work is based on the ET, and they unitarize all the helicity amplitudes using the ET calculation, valid only to describe the longitudinally polarized gauge bosons at high energies. Thus, given this difference between their method and ours, we consider pertinent to make some comments about the discrepancies we have found.

Our predictions match those of the Wizard group for all the $LL\to LL$ amplitudes we have considered, i.e., for all the studied energies and values of the chiral parameters. However, there are some regions of the parameter space in which the cross sections of the other helicity channels differ. In the case in which the purely longitudinal scattering dominates at high energies, both procedures give rise to the same values for the cross sections. If other helicity channels have important contributions to the total cross section, we obtain different predictions. This can be the case if the values of $a_4$ and $a_5$ are very small, of the order of, for instance, $10^{-4}$. The authors in~\cite{Kilian:2014zja,Alboteanu:2008my} themselves comment on the limitations of their approach in this regime, so we are proposing here a way to avoid these limitations.

 
We have seen that different unitarization methods lead to very different predictions for the values of the cross section of the elastic WZ scattering. For this reason one can expect that the translation of these results to the LHC scenario would also show the different behaviors present at the subprocess level. Precisely because of this, the experimental measurements and constraints interpreted using one method or another will be different, and this difference can be understood as a theoretical uncertainty which is precisely the one that we want to quantify in this work. Thus, in the next section, we will present our results for the LHC, and we will give an estimate of this uncertainty in the experimental determination of $a_4$ and $a_5$ due to the unitarization scheme choice.

%
 
 \section{Parameter determination uncertainties at the LHC due to unitarization scheme choice}
 \label{LHC}
 
 In the previous section we learnt that the predictions for WZ scattering observables computed in the EChL framework can be very different depending on the unitarization method we apply to them.  This was manifest at the subprocess level, but now we want to study and quantify these deviations as they would be seen at the LHC. 
 In order to compute the total cross section at the LHC we have first used the simple tool provided by the Effective W Approximation ~\cite{Dawson:1984gx,Johnson:1987tj} and then we have compared this approximate result with the full result from MG5~\cite{Alwall:2014hca,Frederix:2018nkq}. The EWA is the translation to the massive EW gauge bosons case of the familiar Weisz\"{a}cker- Williams, or EPA,  approximation for photons \cite{vonWeizsacker:1934nji,Brodsky:1971ud}. This framework has two important advantages: the first one is that has the intuitive physical interpretation of the distribution functions of the W and the Z as the PDFs in the parton model, and the second one is that it is  computationally simple and, as we will see next, leads to very good results within the context we are working on here. 
 The EWA provides  probability  functions, $f_{W,Z}(x)$, for the W and the Z that describe the probability of the EW gauge boson to be radiated collinearly from a fermion carrying a  fraction $x$ of its total momentum. In order to get the total cross section at the LHC for the full process that starts with protons, these functions, taking quarks as the mentioned fermions, are then convoluted with the PDFs of the quarks\footnote{For this comparison as well as for all the results presented in this work for the LHC, we use the CT10 set of PDFs~\cite{Lai:2010vv}, evaluated at $Q^2=M_W^2$. We utilise the same PDF set for the MG5 comparison, evaluated at the same factorization scale.} and with the corresponding subprocess cross section for the scattering of on-shell EW gauge bosons, $\sigma(WZ \to WZ)$ in our present case. Furthermore, the computation with the EWA requires to separate the different polarizacions for the EW gauge bosons, and to use accordingly the corresponding probability function for the polarized $W$ or $Z$.  For the numerical computation of the cross section at the LHC  with the EWA in this work we have developed our own private PYTHON code. 
 
There are several studies in the literature that use the EWA to obtain reliable estimates. However, not all of them employ the same  probability functions. For this work, we have considered and compared four of these implementations of the EWA. These four implementations are: 1) the original EWA functions  given in \cite{Dawson:1984gx}, including first the  Leading Log Approximation (LLA) ones (eqs. 2.19 and 2.29 in \cite{Dawson:1984gx}); and second 2) the improved ones which go beyond the LLA by keeping ${\cal O}(M_V^2/E^ 2)$ corrections, with $M_V$ the EW gauge boson mass and $E$ the energy of the initial quark (eqs. 2.18 and 2.28 in  \cite{Dawson:1984gx});  3) the EWA functions derived from \cite{Johnson:1987tj}; and 4) the simplified functions  of the beyond LLA given in \cite{Alboteanu:2008my}. In principle, all should lead to similar results for the $pp\to$WZ+X process, and they do at high invariant masses of the final diboson system. Nevertheless, they differ quite a lot at lower energies. It is worth mentioning that to compute the  $pp\to$WZ+X rates with the EWA, one has to consider the contributions from two different subprocesses: the intermediate state with a W and a Z radiated from the initial protons that then scatter, and, in addition, the case in which a W and a photon are radiated and then scatter. The latter is of great importance in the low energy region where it dominates indeed over the other one. For the photon case we have used the well established probability function of the Weisz\"{a}cker- Williams approximation~\cite{vonWeizsacker:1934nji,Brodsky:1971ud}. In order to select the most accurate  probability function for the EW gauge boson case among the ones available in the literature, we have compared the results of the above four mentioned approaches  to the full results for the complete process $pp\to$WZ+X  obtained using MG5. Notice that for this comparison  we have generated MG5 events of the exclusive process $pp\to$WZ$jj$, which automatically contain all the topologies, i.e., the VBS topologies and all the others contributing to the same order in perturbation theory. Besides, in order to compare properly both results, the MG5 one and the EWA one, one has to set particular kinematical cuts on the final state particles.  In particular, as it is well known, in order to regularize  the Coulomb singularity produced by the diagrams with a photon interchanged in a  t-channel, some minimal cuts have to be imposed on the final particles.  Concretely, for this quantitative comparison of the total cross sections we give the following cuts on the transverse momentum and pseudorapidity of the final gauge bosons V and jets $j$, and the angular separation among the jets:
\begin{align}
|p_{T_V}|&> 20 ~{\rm GeV};~~|\eta_V|<2\,;\nn\\
|p_{T_j}|&> 5 ~{\rm GeV};~~|\eta_j|<10\,;~~\Delta R_{jj}>0.1\,,\label{cuts}
\end{align}
both in the EWA and MG5 for the cuts concerning the gauge bosons, and in MG5 events only for the ones concerning the extra jets. 

\begin{figure}[t!]
\begin{center}
\includegraphics[width=0.6\textwidth]{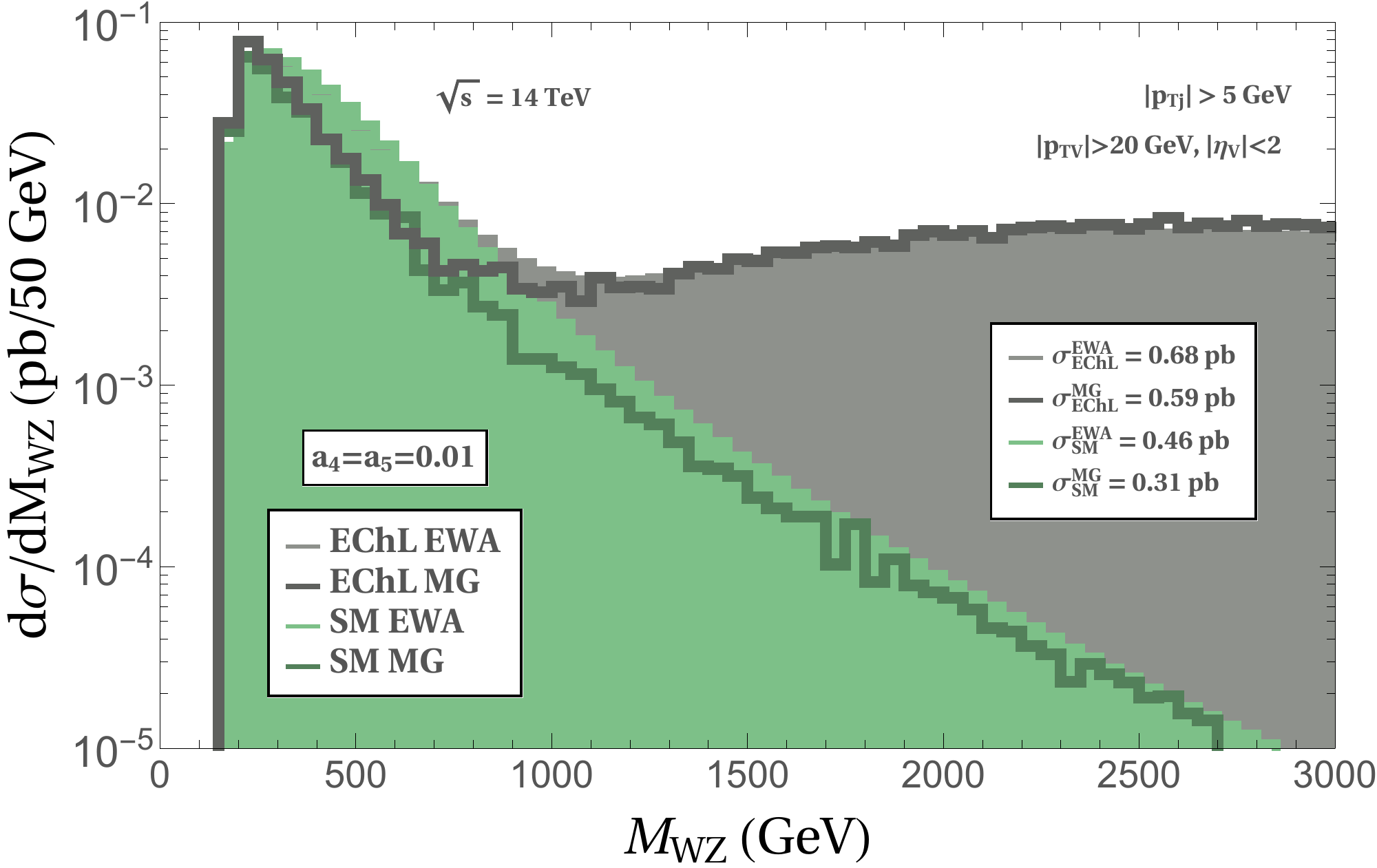}
\caption{Predictions of the differential unpolarized cross section of the process $pp\to$ WZ+X as a function of the invariant mass of the final WZ computed with the EWA (eqs. 2.18 and 2.28 in  \cite{Dawson:1984gx}). SM values (green) and EChL values for $a_4=a_5=0.01$ (gray) are shown. The other chiral parameters are set to their SM value. The MadGraph prediction of $pp\to $WZ$jj$ events with $|p_{T_j}|>5$ GeV is included as a solid line of each corresponding color for reference. All predictions are computed applying the cuts in \eqref{cuts}  and at $\sqrt{s}=14$ TeV.}
\label{fig:EWAcheck}
\end{center}
\end{figure}

With these considerations in mind, we have compared quantitatively the predictions of the $pp\to$WZ+X processes in the SM and in the EChL for $a_4=a_5=0.01$ within the EWA for the four  probability functions considered against the MG5 computation of the $pp\to$WZ$jj$ events. From our numerical comparison (not included here for shortness) we have reached the conclusion that the original,  improved probability functions given in  \cite{Dawson:1984gx} are the ones that better match the MG5 prediction. The others overestimate the probability of radiating a EW gauge boson at low fractions of momentum of the initial quarks, thus missing the correct prediction of the cross section at low energies where most of events lie. In \figref{fig:EWAcheck}, we display the results of the differential cross section distribution with respect to the invariant mass of the final gauge bosons, computed in the SM (green) and in the EChL (gray) for $a_4=a_5=0.01$ using the EWA and employing these  improved probability functions. We also show the MG5 prediction for these same distributions as solid, darker lines of each corresponding color, as well as the total cross sections obtained with both procedures. Cuts in \eqref{cuts} have been required, if applicable, and center of mass energy has been set to $\sqrt{s}=14$ TeV, as it will be considered for the rest of the work. Regarding the comparison shown in this figure, it is manifest that the EWA works remarkably well, specially at high invariant masses. Not only the total MG5 cross section is recovered within a factor 1.5 at the worst in the SM case and 1.15 in the EChL case, but also the invariant mass distributions match considerably well. 

Now that we have checked that our computations obtained with the EWA employing the  improved probability functions provide reliable predictions of $pp\to$WZ+X observables, we move on to characterize the behavior of the different unitarization methods at the LHC. To that purpose, we have convoluted the subprocess cross sections of each of the studied unitarization methods, corresponding to the different curves in \figref{fig:methodssubprocess}, with the EW gauge bosons probability functions and with the CT10 set of PDFs  \cite{Lai:2010vv}, evaluated at $Q^2=M_W^2$.

\begin{figure}[t!]
\begin{center}
\includegraphics[width=0.49\textwidth]{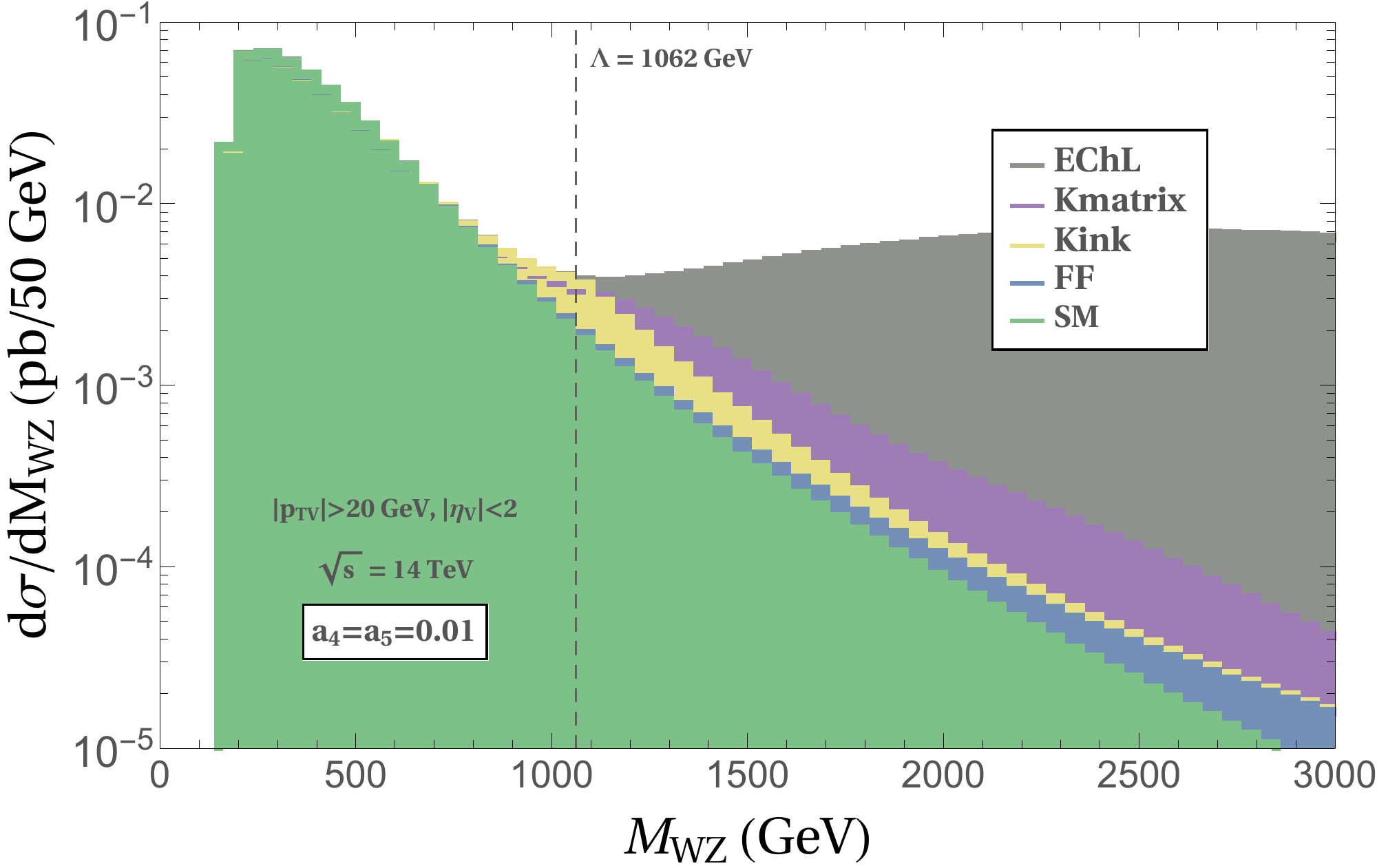}
\includegraphics[width=0.49\textwidth]{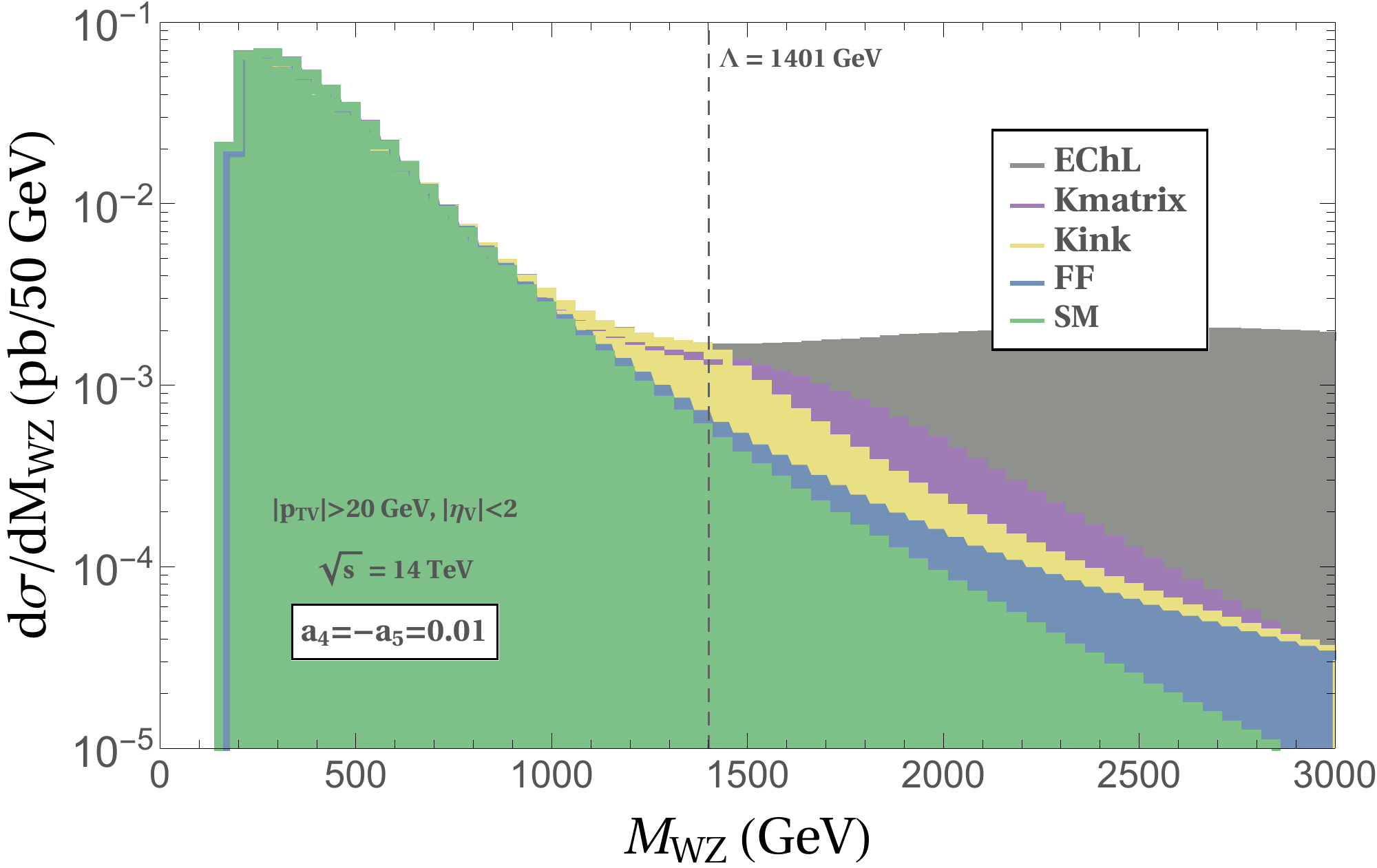}
\caption{Predictions of the differential cross section of the process $pp\to $WZ+X as a function of the invariant mass of the final WZ pair computed with different unitarization methods using the EWA. Predictions of not-unitarized EChL (gray), K matrix unitarization (purple), Kink (yellow), Form Factor (FF, blue) and the SM (green) are shown for two referencial values of the relevant chiral parameters $a_4=a_5=0.01$ (left) and $a_4=-a_5=0.01$ (right). The other chiral parameters are set to their SM value. The unitarity violation scale is also displayed for each case. Predictions are given for $|\eta_V| < 2$ and $|p_{T_V}|>20$ GeV and at $\sqrt{s}=14$ TeV. }
\label{fig:methodsprocess}
\end{center}
\end{figure}

The results are displayed in \figref{fig:methodsprocess}, where we present the invariant mass distributions of the differential cross section of the process $pp\to $WZ+X computed with the EChL for $a_4=a_5=0.01$ (left) and $a_4=-a_5=0.01$ (right) and unitarized with the diverse procedures we have described in the previous section. The non-unitarized EChL and the SM predictions are also shown, for comparison. The unitarity violation scale is marked with a dashed line in each case. The final gauge bosons are required to have $|\eta_V| < 2$ and $|p_{T_V}|>20$ GeV and the evaluation is performed at $\sqrt{s}=14$ TeV. From these curves we can see that the translation of the subprocess results to the LHC is direct, and the conclusions regarding the results are very similar. The different predictions among the various unitarization methods are still manifest, which clearly indicates that the experimental constraints imposed on the EChL parameters will strongly depend on the unitarization method used to analyze the data. Besides, the same pattern of the predictions concerning the relative sign of the chiral parameters is encountered: in the EChL and the K-matrix case, same sign $a_4$ and $a_5$ lead to larger predictions than in the opposite sign case. For the Form Factor and the Kink, the reverse setup is recovered. This still points towards the fact that same sign values of $a_4$ and $a_5$ will be more constrained in the EChL and the K-matrix case, opposite to the Form Factor and the Kink case. The IAM is not shown in these plots since, as we mentioned, it is more suitable for the resonant case. Besides, as we have seen before, in the present non-resonant case, for the chosen particular channel $WZ \to WZ$, and with the simplified 
setup of just two non-vanishing chiral coefficients, $a_4$ and $a_5$, the IAM predictions are very close to the SM ones. Notice that it will not be the case if other channels were considered (for instance, we have checked this explicitly for $WW \to ZZ$) and other chiral coefficients (in particular, we have checked this for $a=0.9$, and $|a_4|,|a_5| \sim {\cal O}(10^{-3}-10^{-4})$) were also non-vanishing.
Regarding the Cut off, it is clear that integrating only up to the unitarity violation scale to obtain the total cross section will lead to much smaller predictions than in the rest of the cases. Finally, it is worth commenting that, as it should be, again all predictions match the EChL one at low invariant masses.

We have now characterized the different predictions of the studied unitarization methods at the LHC. The next step should be to translate these predictions into uncertainties in the extracted constraints on the parameter space of the EChL. In order to do that, we will base our approach upon the ATLAS results for $\sqrt{s}=8$ TeV given in ref.~\cite{Aaboud:2016uuk}. In the mentioned reference, a very sophisticated experimental analysis is performed, specially regarding triggers, background estimations, and event selection. Then, using the K-matrix (or T-matrix) unitarization prescription proposed in~\cite{Kilian:2014zja,Alboteanu:2008my} , the 95\% C.L. exclusion regions in the $[a_4,a_5]$ (sometimes called $[\alpha_4,\alpha_5]$ in the literature) parameter space are obtained. It is beyond the scope of this work to reproduce accurately the experimental analysis of the ATLAS searches. However, there is a consistent way in which we can use their results to obtain the experimental constraints corresponding to other unitarization methods apart from the K-matrix one.

Our approach is the following: first, we take the $a_4$ and $a_5$ values lying on the contour of the WZ observed ``elipse'' provided by the ATLAS study. With those values, we evaluate the total cross section following our K-matrix unitarization procedure for the LHC case, that is, indeed, constant over the mentioned values. This should be equivalent to what ATLAS has performed, since we have checked that for these values of the parameters our prescription matches the one given by the Wizard group. The cross section that we obtain represents the equivalent cross section in our framework to the one that ATLAS has measured experimentally. It is, so to say, a translation between the experimental results and our naive results. Now, what we do is to find the values of $a_4$ and $a_5$ that lead to the same cross section for the other unitarization methods considered in the present work. In this way, we construct the 95\% exclusion regions in the $[a_4,a_5]$ plane for the various unitarization schemes presented in the previous section, to see how they differ in magnitude and shape. By applying this procedure, we are assuming that the selection cuts required to be fulfilled by the ATLAS search affect all our predictions equivalently. This could not be the case, but we expect the differences to be small, so our prescription should be a good first approximation to the issue. Furthermore, it is worth commenting that, regarding the backgrounds, since they are the same to all of our signals, it is well justified to proceed in this way.

 \begin{figure}[t!]
\begin{center}
\includegraphics[width=0.8\textwidth]{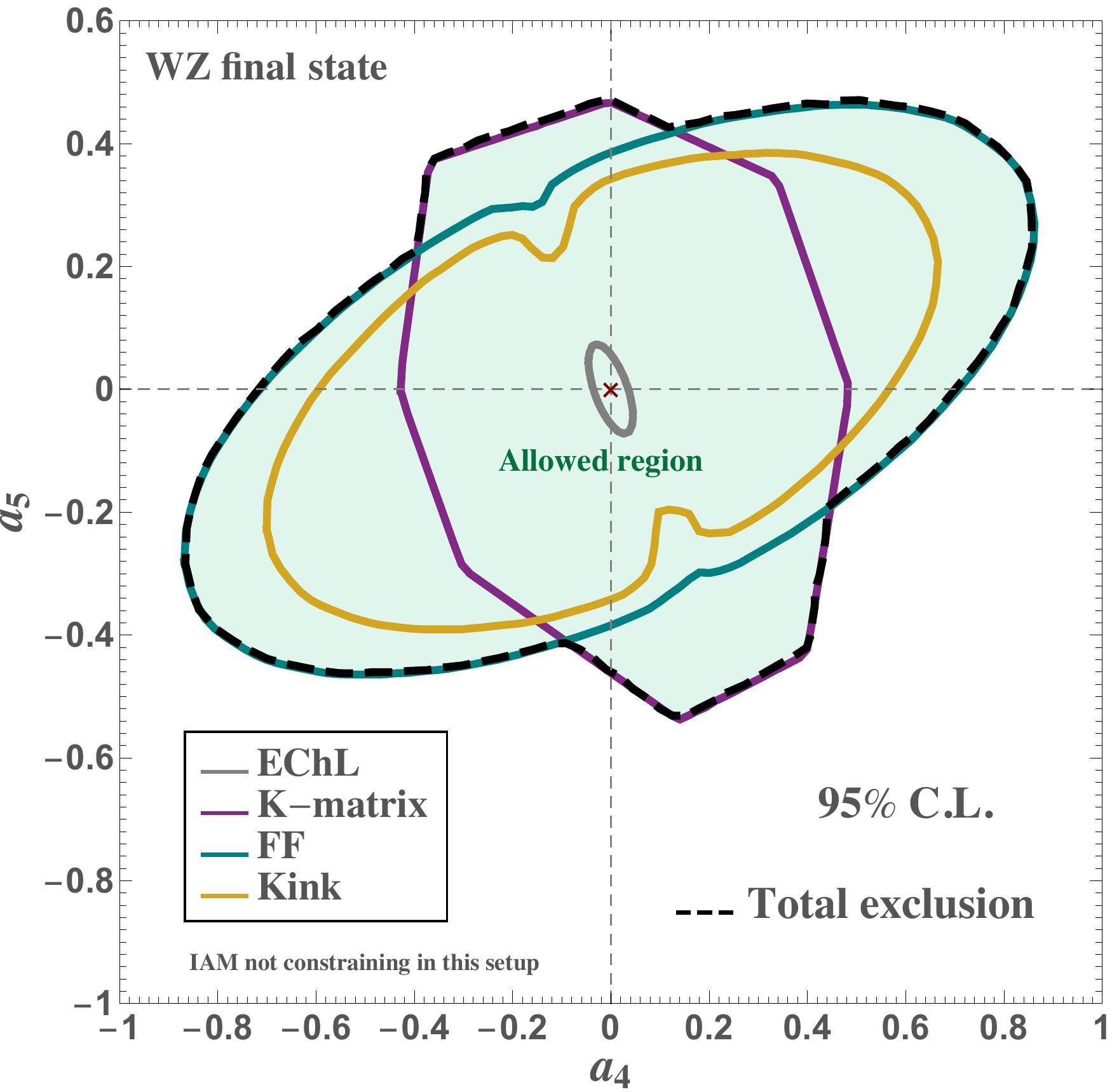}
\caption{Predictions of the 95\% confidence level exclusion regions in the $[a_4,a_5]$ plane for the WZ final state at the LHC with $\sqrt{s}=8$ TeV,  and for the not-unitarized EChL (gray) and the different unitarization methods described in the text: K matrix (purple, corresponding to the solid cyan line in \figref{fig:elipseATLAS}), Kink (yellow) and Form Factor (FF, blue). The total overall exclusion region is the one outside the boundary denoted with a dashed black line. The SM point is marked with a red cross. To obtain this figure we have used the WZ results in \figref{fig:elipseATLAS} as reference.}
\label{fig:elipse}
\end{center}
\end{figure}

The final results of the present work, i.e., the 95\% C.L. exclusion regions in the $[a_4,a_5]$ plane for different unitarization scheme choices, are presented in \figref{fig:elipse}. There we show the corresponding limits for the case in which no unitarization is performed at all (EChL, in gray), for the K-matrix unitarization, matching, of course, the ATLAS results (purple), for the Form Factor prediction (blue) and for the Kink (yellow). We also show the total exclusion region, obtained by the overlap of the former ones.

Many interesting features can be extracted from this figure. First of all, and most importantly, it is indeed very clear that using one unitarization method at a time to interpret experimental data does not consider the full EFT picture. Since there are many unitarization prescriptions that lead to very different constraints, one should take them all into account in order to provide a reliable bound on the EFT parameters. These different constraints can  vary even in an order of magnitude, as it is manifest in \figref{fig:elipse}. For instance, the Form Factor prescription leads to bounds on $[a_4,a_5]$ of the order of [0.8,0.4], roughly speaking, whereas the case in which there is no unitarization performed leads to constraints of the order of [0.04,0.08]. Notice that these latter bounds, i.e., those obtained from the raw, non-unitarized EChL, are not directly comparable to those given in \cite{Sirunyan:2019ksz}, since our results correspond to $\sqrt{s}=8$ TeV and the ones reported in the mentioned reference to $\sqrt{s}=13$ TeV. We leave the precise computation of the 13 TeV results for a future work.
It is also obvious from this figure that the Kink method leads to more stringent constraints than the FF method (the corresponding pseudo-ellipse is smaller and oriented similarly to the FF one). Also the K-matrix method leads to more stringent constraints than the FF one and, in this case, with a different orientation of the pseudo-ellipse (indeed, similar to the EChL one). Interestingly, for the present studied case of non-resonant  $pp \to$ WZ +X events, there is not just a difference in the magnitudes of the bounds of the EChL parameters, but also in the role of each of them, $a_4$ and $a_5$. This feature was already stated before, since in \figref{fig:methodsprocess} we had already seen that points lying in the region of the plane in which $a_4$ and $a_5$ have the same sign should be more constrained in the EChL and the K-matrix case, just in the opposite direction to the Form Factor and the Kink case.

At this point, two further comments have to be made. 
The first concerns the IAM, whose prediction is not present in this figure. It is due to the same argument we have been commenting throughout the text, that can be summarized in the fact that, for our particular setup (non-resonant case with deviation with respect to the SM coming only from the two considered $\mO(p^4)$ operators, i.e., for $a=b=1$ and just $a_4$ and $a_5$ non-vanishing) this method is not suitable to impose reliable constraints on the EChL parameters. Nevertheless, the IAM can be extremely useful when looking for new physics signals at the LHC in the resonant case, as it has been studied in~\cite{Delgado:2017cls}.
The second concerns the Cut off, also not present in the figure. Since this procedure implies to sum events only up a to a determined invariant mass of the diboson system to obtain the total cross section, a problem arises concerning the backgrounds. In our approach, we are always integrating over the whole studied energy region, for all the unitarization method predictions. This means that the background is considered to be the same for all of our signals and we can use the translation from the ATLAS results safely. However, if we now change the picture and integrate over a smaller invariant mass region, such as in the Cut off procedure, we should take into account this same integration over the background, and the pure translation form the ATLAS results fails, since we don't know the background scaling with energy. For this reason, we have not included the Cut off prediction in our final results, but we really do believe that it should be also considered in proper experimental searches. 
 
\section{Conclusions}
\label{Conclusions}

It is undoubtable that effective field theories constitute a remarkable, model independent tool to help us understand the true nature of the electroweak symmetry breaking sector. They typically suffer, however, from unitarity violation problems, due to the energy structure of the operators they contain. If the predictions of observables in such theories violate unitarity from some energy scale upwards, they are, in principle, not compatible with the underlying quantum field theory. Therefore, reliable, unitary predictions are needed to interpret experimental data in order to obtain information about the effective theory and thus about the  dynamics it describes. With the aim of obtaining these predictions, unitarization methods are addressed. There are, nevertheless, many available options to drive non-unitary observables computed with the raw effective theory into unitary ones. This ambiguity supposes then a theoretical uncertainty that has to be taken into account when constraining the parameter space of such theories, that, up to now, has been made using one of these prescriptions at a time or no prescription at all. In this work, we provide a first approximation to quantifying this uncertainty by studying the non-resonant case of the elastic WZ scattering at the LHC.

In order to do so, we use the electroweak chiral Lagrangian or Higgs effective field theory, that is the most approriate EFT if one assumes a strongly interacting EWSB system. With this EChL, we study the violation of unitarity in the WZ$\to$WZ scattering and select the most relevant operators concerning it. These correspond to the ones controlled by the chiral coefficients 
 $a_4$ and $a_5$, which parametrize in this context 
 the anomalous
 interactions among four massive electroweak gauge bosons. Furthermore, at this point, we also analyze the relevance in the unitarization procedure of each of the helicity channels participating in the scattering. Although nowadays most unitarization studies assume that the purely longitudinal scattering is sufficient to understand the unitarity violation processes, we consider of much importance to take into account the whole coupled system of helicity states, since the unitarity conditions relates them. Therefore, we implement this coupled system analysis in our final results. 

Once we have characterized our framework, we study the predictions of different unitarization methods at the subprocess level. We choose five of these methods: Cut off, Form Factor, Kink, K-matrix and Inverse Amplitude Method, which are the most used ones in the literature, to illustrate how different their predictions can be. Their concrete implementation as well as a brief explanation of each of them can be found in Section \ref{Unitarity}. When analyzing each of these methods' predictions at the subprocess level, compared to those of the raw effective theory and of the SM, one is convinced that they lead to very different results, and that this fact should be taken into account at the time to impose constraints on the parameter space of the effective theory.

Moving on to the LHC case, we use the Effective W Approximation to give estimates of the predictions of the various unitarization methods considered for the $pp\to$ WZ+X process. In order to be sure that the EWA works properly for our purpose here, we have first compared its predictions at the LHC of the total cross sections and differential cross sections with the invariant mass $M_{WZ}$,  both in the SM and in the EChL case, with the corresponding full predictions provided by MadGraph. In this comparison, we study various probability functions available in the literature for the massive electroweak gauge bosons and select the ones that better reproduce the MadGraph simulation of the total $pp\to$ WZ+X process. Concretely, we find that the improved EWA functions in \cite{Dawson:1984gx} are the most accurate providing predictions which are in very good agreement with the MadGraph result.  Afterwards, we employ these most accurate EWA functions to obtain the predictions of the invariant mass $M_{ZW}$ distributions of the differential cross section of the $pp\to$ WZ+X events for the different unitarization methods discussed in this work. We conclude again that the various unitarization methods provide very different predictions not only for the subprocess but also for the total process at the LHC. 

Finally, we construct, based on the ATLAS results for $\sqrt{s}=8$ TeV given in~\cite{Aaboud:2016uuk}, the 95\% exclusion regions in the $[a_4,a_5]$ plane for the various unitarization schemes. The main results of the work are contained in \figref{fig:elipse}, from which very interesting features can be extracted. The most important of them is that it is indeed very clear that using one unitarization method at a time to interpret experimental data does not consider the full effective theory picture. Since there are many unitarization prescriptions that lead to very different constraints, one should take them all into account in order to provide a reliable bound on the EFT parameters. These different constraints can vary even in an order of magnitude. As an example, the Form Factor method leads to bounds on $[a_4,a_5]$ of the order of $\sim[0.8,0.4]$ whereas the pure EChL prediction, without unitarization, leads to constraints of the order of $\sim[0.04,0.08]$. Furthermore, the differences do not lie just in the magnitudes of the bounds, but in the role of $a_4$ and $a_5$, what can be seen in the shapes of the different exclusion regions. 

The main conclusion of this work, is, therefore, that there is a theoretical uncertainty present in the experimental determination of effective theory parameters due to the unitarization scheme choice. A first approximation to this uncertainty has been quantified in the present work analyzing the predictions of $pp\to$ WZ+X events at the LHC from the EChL in terms of $a_4$ and $a_5$ and with different unitarization methods. We believe that it is important to take these uncertainties into account when relying upon experimental values of the constraints of effective theory parameters, in order to consider the full effective theory properties correctly.


\section*{Acknowledgments}
We would like to thank warmly the various clarifications provided by Antonio Dobado on the unitarization procedure of the coupled helicity amplitudes system which were of great help in the early stages of this work. We also warmly thank the interesting and clarifying comments by Enrique Fernandez-Martinez about the statistical analysis when producing the final figure.  C.G.G wishes to thank Xabier Marcano and Javier Quilis for their invaluable help during the process of obtaining the final results of the work.
This work is supported by the European Union through the ITN ELUSIVES H2020-MSCA-ITN-2015//674896 and the RISE INVISIBLESPLUS H2020-MSCA-RISE-2015//690575, by the CICYT through the projects FPA2016-78645-P, by the Spanish Consolider-Ingenio 2010 Programme CPAN (CSD2007-00042) and by the Spanish MINECO's ``Centro de Excelencia Severo Ochoa''  Programme under grant SEV-2016-0597.

%
\bibliographystyle{JHEP}
\bibliography{GHM}
\end{document}